\documentclass[compsoc,conference,a4paper,10pt,times]{IEEEtran}
\IEEEoverridecommandlockouts

\usepackage{tikz}
\usepackage{amsmath}
\usepackage[htt]{hyphenat}

\usepackage{graphicx}
\usepackage{subcaption}
\usepackage{algorithm}
\usepackage{listings}
\usepackage{gensymb}
\usepackage[utf8]{inputenc}
\DeclareUnicodeCharacter{2212}{-}
\usepackage{mathtools}
\usepackage{algpseudocode}
\usepackage{multirow}
\usepackage{amssymb}
\usepackage{url}
\usepackage{hyperref}
\usepackage{xcolor,colortbl}
\usepackage{caption}    % To create space between listing caption and listing
\usepackage{booktabs}
\usepackage{pifont}
\usepackage{verbatim}
\usepackage{tikz}

\usepackage{makecell}

\usepackage{amsmath, amssymb, amsfonts}
\usepackage{enumitem}
\usepackage{xspace}
\usepackage{listings}
\usepackage{xcolor}
\usepackage{booktabs}
\usepackage{tabularx}
\usepackage{makecell}
\usepackage{amssymb}
\usepackage{pifont}

\usepackage{rotating}

\definecolor{darkpastelgreen}{rgb}{0.01, 0.75, 0.24}
\definecolor{darkpastelred}{rgb}{0.76, 0.23, 0.13}
\definecolor{red}{HTML}{EA6B66}
\definecolor{green}{HTML}{97D077}
\definecolor{blue}{HTML}{7EA6E0}

\usepackage{textcomp}
\usepackage{pifont}% http://ctan.org/pkg/pifont
\newcommand{\attack}{\textsc{LeapFrog}\xspace}%
\renewcommand{\paragraph}[1]{\medskip \noindent {\bf #1.}}
\newcommand{\hl}[1]{\textcolor{black}{#1}}

\lstdefinestyle{customc}{
  backgroundcolor=\color{white},
  basicstyle=\footnotesize\ttfamily,
  columns=fullflexible,
  breakatwhitespace=false,      
  breaklines=true,                
  captionpos=b,                    
  commentstyle=\color{mGreen}, 
  extendedchars=true,              
  frame=single,                   
  keepspaces=true,             
  keywordstyle=\color{blue},      
  language=c++,                 
  numbers=left,                
  numbersep=5pt,                   
  numberstyle=\tiny\color{mGray}, 
  rulecolor=\color{mGray},        
  showspaces=false,               
  showtabs=false,                 
  stepnumber=1,                  
  stringstyle=\color{magenta},    
  tabsize=1,                      
  title=\lstname,
  morecomment=[f][\lstbg{red!20}]-,
  morecomment=[f][\lstbg{green!20}]+,
  morecomment=[f][\textit]{@@},
}

\lstdefinestyle{customc}{
  belowcaptionskip=1\baselineskip,
  moredelim=**[is][\color{red}]{@}{@},
  breaklines=true,
  frame=L,
  numbers=left,
  xleftmargin=13pt,
  numbersep=5pt, % Space between line numbers and code
  language=C++,
  showstringspaces=false,
  basicstyle=\footnotesize\ttfamily,
  keywordstyle=[1]\bfseries\color{blue!40!black},
  commentstyle=\itshape\color{purple!40!black},
  identifierstyle=\color{black},
  stringstyle=\color{orange},  
  keywords=[2]{matched, authenticated, result, ret, fast\_auth\_result},
  keywordstyle=[2]\bfseries\color{darkpastelred},
  morekeywords={munmap},
}
\newcommand*\blackcircled[1]{\tikz[baseline=(char.base)]{
            \node[shape=circle, draw, inner sep=1pt, fill=black, text=white, font=\sffamily\scriptsize] (char) {#1};}}

\def\BibTeX{{\rm B\kern-.05em{\sc i\kern-.025em b}\kern-.08em
    T\kern-.1667em\lower.7ex\hbox{E}\kern-.125emX}}

\usepackage{fancyhdr}
\fancypagestyle{firstpage}{
  \fancyhf{} % clear all header and footer fields
  
  \fancyfoot[C]{%
    \parbox{\textwidth}{%
      \centering % This centers the content within the parbox
      Approved for Public Release; Distribution Unlimited. Public Release Case Number 25-1258 \\
      ©2025 The MITRE Corporation. ALL RIGHTS RESERVED. - \thepage%
    }%
  }
}

\begin{document}

% Your document content here

%\clearpage % End of the first page
%\restoregeometry % Restore the original geometry for the rest of the document
%-------------------------------------------------------------------------------

\newcommand{\tool}{\textit{MFS}\xspace}

\title{\attack: The Rowhammer Instruction Skip Attack}

%\author{\IEEEauthorblockN{Anonymous Author}
%\IEEEauthorblockA{Institution\\
%Georgia Institute of Technology\\
%Atlanta, Georgia 30332--0250\\
%Email: http://www.michaelshell.org/contact.html}
%}}

%for single author (just remove % characters)
% \author{Anonymous Author}

\author{
\begin{tabular}{ccc}
    {\rm Andrew Adiletta} & {\rm M. Caner Tol} & {\rm Kemal Derya} \\
    \phantom{MMMMM} MITRE \phantom{MMMMMM} & Worcester Polytechnic Institute & Worcester Polytechnic Institute \\
    aadiletta@mitre.org & mtol@wpi.edu & kderya@wpi.edu \\
\end{tabular}
\\[15pt] % Adjust spacing between rows as needed
\begin{tabular}{cc}
    {\rm Berk Sunar} & {\rm Saad Islam} \\
    Worcester Polytechnic Institute & Worcester Polytechnic Institute \\
    sunar@wpi.edu & saad.islam@fulbrightmail.org \\
\end{tabular}
}

\maketitle

\thispagestyle{firstpage}

%\pagestyle{fancy}

%%%%%%%% COMMENT OUT BEFORE SUBMISSION TO CCS%%%%%%%%%%%%
% \pagestyle{plain}
%%%%%%%%%%%%%%%%%%%%%%%%%%%%%%%%%%%%%%%%%%%%%%%%%%%%%%%%
%%%%%%%%%%%%%%%%%%%%%%%%%%%%%%%%%%%%%%%%%%%%%%%%%%%%%%%
%%%%%%%%%%%%%%%%%%%%%%%%%%%%%%%%%%%%%%%%%%%%%%%%%%%%%%%
\begin{abstract}
Since its inception, Rowhammer exploits have rapidly evolved into increasingly sophisticated threats compromising data integrity and the control flow integrity of victim processes. Nevertheless, it remains a challenge for an attacker to identify vulnerable targets (i.e., Rowhammer gadgets), understand the outcome of the attempted fault, and formulate an attack that yields useful results. 

In this paper, we present a new type of Rowhammer gadget, called a LeapFrog gadget, which, when present in the victim code, allows an adversary to subvert code execution to bypass a critical piece of code (e.g., authentication check logic, encryption rounds, padding in security protocols). The LeapFrog gadget manifests when the victim code stores the Program Counter (PC) value in the user or kernel stack (e.g., a return address during a function call) which, when tampered with, repositions the return address to a location that bypasses a security-critical code pattern. 

This research also presents a systematic process to identify LeapFrog gadgets. This methodology enables the automated detection of susceptible targets and the determination of optimal attack parameters. We first show the attack on a decision tree algorithm to show the potential implications. Secondly, we employ the attack on OpenSSL to bypass the encryption and reveal the plaintext. We then use our tools to scan the Open Quantum Safe library and report on the number of LeapFrog gadgets in the code. Lastly, we demonstrate this new attack vector through a practical demonstration in a client/server TLS handshake scenario, successfully inducing an instruction skip in a client application. Our findings extend the impact of Rowhammer attacks on control flow and contribute to developing more robust defenses against these increasingly sophisticated threats.
\end{abstract}
% \begin{IEEEkeywords}
% Kernel Stack, Program Counters, Instruction Skip
% \end{IEEEkeywords}
%%%%%%%%%%%%%%%%%%%%%%%%%%%%%%%%%%%%%%%%%%%%%%%%%%%%%%%
%%%%%%%%%%%%%%%%%%%%%%%%%%%%%%%%%%%%%%%%%%%%%%%%%%%%%%%
%\section{Introduction}
%%%%%%%%%%%%%%%%%%%%%%%%%%%%%%%%%%%%%%%%%%%%%%%%%%%%%%%
% \begin{enumerate}
% \item Take POC and dump stack
% \item Annotate ASM with vaddr 
% \item Craft vulnerable target
% \item Locate PC in stack
% \item Simulate flip in memory
% \item Attack PC with Rowhammer
% \item Find Real World Target
% \end{enumerate}

%%%%%%%%%%%%%%%%%%%%%%%%%%%%%%%%%%%%%%%%%%%%%%%%%%%%%%%
\section{Introduction}\label{sec:introduction}
%%%%%%%%%%%%%%%%%%%%%%%%%%%%%%%%%%%%%%%%%%%%%%%%%%%%%%%

%\smallskip
%\noindent
%{\bf Rowhammer Attacks}
The miniaturization of DRAM technology has inadvertently increased the susceptibility to bit flips and reliability issues. To mitigate data corruption, DRAM rows are refreshed at regular intervals, typically every 64 milliseconds. However, Kim et al. \cite{kim2014flipping} discovered that rapid and repeated access to adjacent rows could accelerate charge leakage, leading to bit flips before the scheduled refresh, a phenomenon known as the Rowhammer effect \cite{kim2014flipping}. Expanding on this, Seaborn et al. \cite{seaborn2015exploiting} demonstrated an even more efficient method known as the double-sided Rowhammer, which exacerbates the issue.

Further developments in exploiting the Rowhammer vulnerability have been numerous. Gruss et al. \cite{gruss2018another} achieved root privileges by flipping opcodes in the \texttt{sudo} binary using a single-location hammering technique. Furthermore, Gruss et al. \cite{2016Rowhammerjs} and Ridder et al. \cite{desmash} demonstrated the feasibility of launching Rowhammer attacks remotely via JavaScript. Tatar et al. \cite{tatar2018thRowhammer} and Lip et al. \cite{lipp2020nethammer} extended the reach of Rowhammer to network-based attacks. Its applicability has also been demonstrated in cloud environments \cite{xiao2016one, cojocar2020we} and on hybrid FPGA-CPU platforms \cite{weissman2019jackhammer}. Importantly, Kwong et al. \cite{kwong2020rambleed} revealed that Rowhammer poses not only an integrity threat but also compromises confidentiality.

Efforts to detect ~\cite{irazoqui2016mascat,chiappetta2016real,zhang2016cloudradar,herath2015these,payer2016hexpads,gruss2016flush+,aweke2016anvil,corbet} and neutralize \cite{2016Rowhammerjs, van2016drammer, brasser2017can} Rowhammer attacks have been substantial. However, Gruss et al. \cite{gruss2018another} demonstrated the ineffectiveness of these countermeasures. Moreover, Cojocar et al. \cite{cojocar2019ecc} questioned the security of ECC as a countermeasure. The Target Row Refresh (TRR) hardware countermeasure was recently also circumvented, as shown by Frigo et al. \cite{frigo2020trrespass} and further exploited by Ridder et al. \cite{desmash} to target DDR4 chips with TRR. In particular, a recent study by Kogler et al. \cite{kogler2022halfdouble} highlighted the feasibility of hammering beyond adjacent locations to bypass TRR defenses.

The software exploits enabled by Rowhammer were further studied in recent research. Tobah et al.~\cite{gogogadget} introduced the notion of \textbf{Rowhammer gadgets} and a specialized attack. Specifically, if a victim code is designed to return benign data to an unprivileged user, and uses nested pointer dereferences, Rowhammer can be used to flip these pointers thereby gaining arbitrary read access in the victim’s address space. Adiletta et al.~\cite{adiletta2023mayhem} demonstrated that even internal CPU elements such as register values, which are occasionally saved to the stack, can be vulnerable to Rowhammer when they are temporarily stored in the stack and flushed to memory. Upon reloading, these corrupted values are returned to the registers, potentially leading to the execution of faulty stack variables and security breaches.

%\smallskip
%\noindent
%{\bf Efforts to detect}

%\smallskip
%\noindent
%{\bf Countermeasures in Crypto Libraries}
The threat of physical fault injection attacks has been acknowledged in the cryptographic community for some time \cite{Boneh2015OnTI}. For instance, \texttt{OpenSSL} incorporated error checks in CRT-based exponentiation early on to combat Bellcore attacks \cite{Boneh2015OnTI}. However, fault injection techniques have successfully compromised Elliptic Curve Parameters in the \texttt{OpenSSL} library \cite{DBLP:conf/eurosp/0002T19}. Similarly, Rowhammer-induced fault attacks in \texttt{WolfSSL}, leading to ECDSA key exposure, were revealed in \cite{mus2023jolt, derya2024fault+}. The vulnerabilities occurred during the TLS handshake process, involving the signing operation with private ECC keys. WolfSSL responded by introducing \texttt{WOLF\_SSL\_CHECK\_SIG\_FAULTS} and \texttt{WOLFSSL\_BLIND\_PRIVATE\_KEY}, a series of checks during the signing stages to detect data tampering \cite{nist_2022, CVE-2024-5288}. 

A recent work by Adiletta et al.~\cite{adiletta2023mayhem} targeted sensitive stack variables via Rowhammer threatening data integrity. In this paper, we instead target the control-flow-integrity (CFI) and subvert the execution flow for malicious ends, e.g. to bypass sensitive sections of code user authentication and data encryption. For this, we introduce \attack\ a new Rowhammer attack vector that targets the PC when stored in the stack during function calls and context switches. Not all PC manipulations will yield useful results, as some jumps within the code will result in errors, like segfaults, or simply will not bypass the intended code logic. To explore the massive attack surface, we introduce an automated tool that dynamically analyzes code to detect this type of Rowhammer gadget~\cite{gogogadget}.

%%%%%%%%%%%%%%%%%%%%%%%%%%%%%%%%%%%%
\subsection{Our contributions}
%%%%%%%%%%%%%%%%%%%%%%%%%%%%%%%%%%%%
We introduce a novel approach for identifying \attack\ Rowhammer gadgets capable of corrupting the PC, utilizing a combination of GDB, the Intel Pintool, and the Linux Process Interface. 

% TODO: Restate contributions to focus on end-to-end TLS handshake, and talk about scanning success of OpenSSL and sudo

Our contributions are fivefold:
\begin{enumerate}
\item We introduce the concept of \attack gadgets, which allows an attacker to bypass security critical areas of code by faulting the PC value stored in stack.
% More clearly spell out the contributions

\item We introduce the first simulation tool designed to identify LeapFrog gadgets. This tool represents an improvement over existing methodologies \cite{YimMethodology2016} by systematically analyzing binaries with our Intel \texttt{Pin}-based tool called \tool and incorporating time-domain analysis in simulations.
\item We scan the Open Quantum Safe library signature scheme, OpenSSL encryption, and a machine learning model - and quantify the number of potential LeapFrog gadgets present in the code. 

\item We validate the feasibility of this attack in practical scenarios by successfully bypassing a TLS handshake in standard OpenSSL implementations. 

\item We propose and evaluate countermeasures against the LeapFrog attack, offering insights into enhancing the resilience of systems against such advanced Rowhammer based exploits.
\end{enumerate}

\subsection{Discussion on Novelty}
\label{sec:discussion-novelty}

Our work extends control-flow attacks by focusing on the saved Program Counter (PC) in the stack. Table~\ref{tab:novelty-compare} summarizes how LeapFrog differs from other techniques. Rowhammer-based attacks on stack variables \cite{adiletta2023mayhem}, opcodes \cite{gruss2018another}, or pointers \cite{gogogadget} do not skip instructions by flipping the PC and have different mitigations. Meanwhile, classical ROP methods \cite{roemer2012return} rely on controlling and overwriting larger portions of the stack, making it potentially detectable by stack canaries. By flipping only the stored PC bits, LeapFrog bypasses many existing defenses. Additionally, LeapFrog requires synchronization like most fault injection attacks, but uniquely provides a dedicated detection framework (\tool) that helps identify vulnerable locations automatically.

\begin{table}[h]
\centering
\scriptsize
\setlength{\tabcolsep}{8.5pt}
\begin{tabular}{lllcc}
\toprule
\textbf{Tool} & \textbf{Target} & \textbf{Mitigation} & \textbf{Sync} & \textbf{Detect} \\
\midrule
\textbf{LeapFrog} & PC & PC Redundancy & \checkmark & \checkmark \\
\cite{adiletta2023mayhem} & Stack Vars & Magic Vars & \checkmark & \ding{55} \\
\cite{gruss2018another} & Opcode & Compiler & \ding{55} & \ding{55} \\
\cite{gogogadget} & Pointers & Fix Code Pattern & \checkmark & \checkmark \\
\cite{derya2024fault+} & Heap Vars & Blinding/Masking & \checkmark & \ding{55} \\
\cite{roemer2012return} & PC & Stack Canaries & \ding{55} & \checkmark \\
\bottomrule
\end{tabular}
\caption{Comparison of LeapFrog with other return-oriented or data-corruption attacks. LeapFrog bypasses typical stack protections by flipping only PC bits. We compare if precise attacker/victim syncronization is required, and it there is an easy-to-use open source tool for detection.}
\label{tab:novelty-compare}
\end{table}

%%%%%%%%%%%%%%%%%%%%%%%%%%%%%%%%%%%%%%%%%%%%%%%%%%%%%%%
\section{Background}\label{sec:background}
%%%%%%%%%%%%%%%%%%%%%%%%%%%%%%%%%%%%%%%%%%%%%%%%%%%%%%%
\smallskip
\noindent
{\bf Rowhammer} DRAM is stored in an array architecture of memory cells, where a capacitor and a transistor form each cell in the array. Each cell, capable of storing a value of either 1 or 0, is connected along word lines that extend across the row. Additionally, bit lines intersect the word lines perpendicularly, linking them to each cell. When the bit lines are brought into opposition (where one goes high and one goes low), positive feedback from a sense amplifier sets the state of the cell to be high or low. The sense amplifier consists of two cross-connected inverters between the cells.  For the cell to retain its state, the sense amplifier must be disconnected \cite{an2021rowhammer}.

The sense amplifiers must be disconnected to read the cell, so the target word-line must be brought high. The charge in the capacitor will bring one of the bit lines high if the cell has a value of 1 during reading, then the sense amplifier will be reconnected, and the row will have the sense amplifier outputs latched \cite{rahman2013design}.

The Rowhammer attack works by abusing the sense amplifier's ability to set the state of the cell. The capacitor will leak voltage and must undergo a refresh of voltage every 64 ms (or less) according to the standard JETEC convention. Reading the cell introduces noise into the system due to fluctuating voltages, and the noise can be amplified by the sense amp. As semiconductor technology improves, transistors shrink in size and the operating voltage also shrinks. The resulting cells are then able to operate at a higher speed, and the noise margins are reduced. Shrinking in size increases the signal coupling across traces and devices and magnifies crosstalk. When combined with the lower operating voltage and sharper edges due to higher speeds, the ratio of the crosstalk to the supply voltage increases significantly as technology improves making it easier for an adversary to exploit this type of attack. These can result in errors being written to the DRAM \cite{an2021rowhammer}.

\smallskip
\noindent
{\bf Instruction Skipping}
Instruction-skipping attacks are a type of fault attack that targets the normal execution flow of a program, particularly in embedded systems and secure circuits. Historically these techniques often needed physical access due to the timing and precision required to skip instructions. For example, laser fault injection has been demonstrated in inducing instruction skips in AES (Advanced Encryption Standard) that resulted in leaking private encryption keys \cite{breier2015laser}. In another example \cite{riviere2015high}, researchers demonstrate how electromagnetic fault injection can effectively induce instruction skipping in the ARMv7-M architecture, specifically targeting AES. 

In terms of countermeasures for instruction skips specifically, \cite{moro2014formal} addresses the vulnerability of embedded processors to instruction skip attacks. The paper acknowledges that while countermeasures based on temporal redundancy have been proposed, they are not entirely effective against double fault injections over extended time intervals. 

\smallskip
\noindent
{\bf Kernel Stack vs User Stack}
The user stack operates in user mode and is operated by user-level processes. Each user process has its own stack that stores local variables, function parameters, return addresses, and the control flow of the program. This stack is limited in size and is specific to the user space process, ensuring isolation and security from other processes.

On the other hand, the kernel stack operates in kernel mode, a privileged mode of operation for the system's kernel. Each thread of a process has its own kernel stack. This stack is used when the process executes system calls or when it is interrupted and the kernel needs to perform operations on behalf of the process. The kernel stack handles system-critical lower-level operations such as interrupt handling, system call implementation, and managing hardware interactions. It is kept separate from the user stack for security and stability, ensuring that user processes cannot directly access or interfere with kernel-level operations. Data stored in the kernel stack includes CPU context for system calls, interrupt state information, and other kernel-specific data, while the user stack holds user-level process data like function calls and local variables. This separation reinforces the security and stability of the operating system by isolating user applications from the core kernel functions.

\smallskip
\noindent
{\bf Program Counters}
Program Counters (PCs), also known as Instruction Pointers, hold the memory address of the next instruction to be executed by the CPU. This mechanism ensures that instructions are executed in the correct sequence.
The value of the Program Counter is typically not stored in the stack; rather, it's stored in a dedicated register within the CPU. During the execution of a program, the PC is automatically incremented after each instruction is fetched, pointing to the subsequent instruction. However, during certain operations like function calls and interrupts, the PC value may be changed abruptly to a new address. In such cases, the return address (the original PC value) is often stored in the stack to enable the program to return to the correct point in the program after the operation is complete. For user function calls, the PC value is pushed to the user stack, but if there is an exception, signal handler, or system call, the PC gets pushed to the kernel stack. This mechanism facilitates the smooth flow of program execution.

\smallskip
\noindent
{\bf Process Degradation} Process degradation in computing refers to the intentional slowing down of a processor to create favorable conditions for certain types of attacks. A notable contribution in this field, HyperDegrade \cite{aldaya2022hyperdegrade}, combines previous approaches \cite{allan2016amplifying} with the use of simultaneous multithreading (SMT) architectures to significantly slow down processor performance, achieving a slowdown that is orders of magnitude greater than previous methods. 
%
%The use of Self-Modifying Code (SMC) in HYPERDEGRADE plays a significant role in enhancing performance degradation. 
It utilizes collateral Self Modifying Code (SMC) events to induce ``machine clears", where the entire CPU pipeline is flushed, resulting in severe performance penalties. This process is triggered by cache line eviction, causing the invalidation of instructions in the victim's L1 instruction cache, which the CPU may interpret as an SMC event. %The resultant pipeline flushes force re-fetching of instructions from memory, increasing L1 cache misses. 
This mechanism amplifies the degradation effect, as instructions are sometimes fetched multiple times, leading to substantial slowdowns in CPU performance.
This slowdown enhances the time granularity for FLUSH+RELOAD \cite{Gullasch2011cachegames} attacks, enabling more effective exploitation of side-channel vulnerabilities in systems. The attack not only explores the implementation of this technique but also investigates the root causes of performance degradation, particularly focusing on cache eviction. Their findings have substantial implications in the realm of cryptography, as evidenced by the amplification of the Raccoon attack \cite{merget2021raccoon} on TLS-DH key exchanges and other protocols.%, demonstrating its potential in practical attack scenarios against OpenSSL vulnerabilities.

%% TODO: Add background section on Rowhammer [DONE]

%%%%%%%%%%%%%%%%%%%%%%%%%%%%%%%%%%%%%%%%%%%%%%%%%%%%%%%

\section{Related Work}
% TODO (ANDREW):

% Shoot down the following papers
% 1. Op Code Flipping
% 2. Go-go gadget
% 3. The Rowhammer methodology [Google]
% 4. Mayhem
% 5. PTE paper
%Similar work \cite{gruss2018another} achieved privilege escalation through opcode flipping. Researchers loaded the \texttt{sudo} binary into memory from user space and flipped a bit in the opcode of the binary such that an incorrect password would cause authentication and the correct password would cause mis-authentication. They mention flipping conditional jumps that change program execution flow. While our work also can result in privilege escalation, we do so by attacking the process during run time and forcing the process to jump to an unexpected line of code by flipping the PC value. 

To attack the binary during runtime, we had to overcome timing challenges as well as different detection problems to find vulnerable areas in the code. In \cite{gruss2018another} researchers used \texttt{mmap} to map the target binary into a vulnerable page in memory, demonstrating how memory waylaying and memory-chasing techniques can force the mapped binary into the target page. This attack can potentially be mitigated by making the process execute only, and thus cannot be mapped with the \texttt{mmap} command. In contrast, our work can attack binaries that are unreadable from userland and are executed only. Additionally, our attack works on fundamentally different mechanics, so targets not susceptible to \cite{gruss2018another} may be susceptible to ours. 

Another related work \cite{gogogadget} demonstrates that code using nested pointer dereferences can corrupt bits in these pointers to reveal data to an unprivileged user. They demonstrate this vulnerability on \texttt{ioctl} given they can flood the kernel heap with data by spawning processes (a method they call ``spraying"), increasing the probability a single bit-flip will point to malicious data in the heap that points to the location of secret data. Our work compliments and improves upon this prior work by increasing the number of vulnerable code patterns since their work relies on the presence of specific code patterns that may not be present in the victim code. 

Lastly, \cite{yim2016rowhammer} demonstrates a Rowhammer attack methodology where researchers emulated Rowhammer bitflips on targets. They introduced the idea of simulating a flip in the \texttt{EIP} register value in the stack, which can force the execution to jump from kernel code to user code, like the ret2usr attack \cite{koruyeh2018spectre}. However, attacks that cause privilege escalation by jumping from kernel code to user code are mitigated by \texttt{SMAP} \cite{lwn_smap}, which prevents the kernel from executing userland instructions. Our attack forces a process to jump within its own code space and privilege space and thus is not affected by \texttt{SMAP} and introduces attack surfaces on new code patterns. 

% Start

\section{Threat Model}

Similar to other Rowhammer attacks we assume the attacker is co-located on the same system as the victim ~\cite{kim2014flipping,gruss2018another,xiao2016one,cojocar2020we,2016Rowhammerjs}. Co-location is a common threat model for many micro-architectural side-channel attacks and fault attacks ~\cite{Lipp2018meltdown,Kocher2018spectre,canella2019fallout,vanbulck2020lvi,vanschaik2019ridl}.  We do not assume root privilege or physical access to the machine. 
%For some attacks, we assume that the attacker can send signals to pause it during the attack. %, such as \texttt{sudo}. 
%However, we do not need to be root to send signals. 
We only assume that the system has TRR enabled, and bypasses TRR with a many-sided attack \cite{frigo2020trrespass}.

%%%%%%%%%%%%%%%%%%%%%%%%%%%%%%%%%%%%%%%%%%%%%%%%%%%%%%%
\section{LeapFrog Attack}\label{sec:LeapFrog_attack}

% TODO: Add a paragraph formalizing LeapFrog gadget

% TODO: Convert C code to Drawio
\begin{comment}
\begin{figure}%[!h]
\begin{lstlisting}[frame=single,
                    language=C++,
                    label={lst:TLS_C},
                    caption= Simple TLS handshake example where a client process is trying to connect to a server and the PC value origin which could be targeted to enable a mis-authentication is 4 bit flips away]
unsigned char sig_buf[72];
int sig_len = 0;
struct receive_return result = 
    wait_receive(client_fd, sig_buf);
...
// Verify the signature
if (verify_message(message, 
    sizeof(message),
    signature, ec_key)==SUCCESS){
    pass = 1;
}
...
if (pass != 0) {
    fprintf(stdout, 
    "Server Authenticated\n");
}
else{
    fprintf(stdout, 
    "Server Not Authenticated\n");
}
\end{lstlisting}
\end{figure}
\end{comment}
% TODO: Turn this into DRAW.IO

\begin{figure}[!b]
    \centering
    \includegraphics[width=\columnwidth]{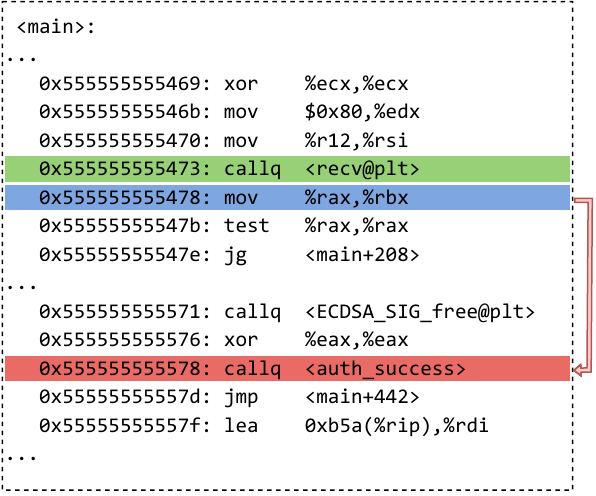}
    \caption{LeapFrog gadget in TLS handshake $addr_{src}$, the PC value that fault is injected into, is highlighted in \colorbox{blue}{blue}. The new value is highlighted in \colorbox{red}{red}. The fault is injected during the execution of the function call highlighted in \colorbox{green}{green}.}
    \label{lst:tls_asm}
\end{figure}

%Central to our investigation is the concept of \textit{LeapFrog gadgets}, an important element in performing Rowhammer-based attacks targeting the Program Counter (PC). 
\textit{LeapFrog} gadgets are exploitable in scenarios where a process undergoes a context switch or executes a function call, leading to the storage of the PC value in either the kernel or user stack. The ingenuity of LeapFrog gadgets lies in their susceptibility to Rowhammer-induced bit flips due to them being stored in DRAM, enabling an attacker to alter the PC value subtly. This manipulation is designed to redirect the execution flow to a different code segment, ideally with minimal bit changes due to the blunt nature of Rowhammer and the higher probability of finding a faulty memory location with few or one faulty bits in the right location. In this paper, we assume that we can successfully find a 1-bit flip within a page that is in the right location to fault the PC value to force the intended instruction to skip.

In Figure \ref{lst:tls_asm} the storage of the PC value occurs in the kernel stack during the execution of \texttt{wait\_receive}. In this scenario, a malicious server can hold the client process at the \texttt{wait\_receive} function while hammering the PC value to force the process to jump to a new location upon returning from the function. In our assembly code analysis in Figure \ref{lst:tls_asm}, we observe the original PC value is an address \texttt{0x555555555478}. Through strategic bit flips, this value can be altered to \texttt{0x555555555578}, effectively enabling an instruction skip (skipping one or more instructions) and jumping from the function call in wait\_receive directly to a later point in the execution, bypassing the critical server authentication check. The practicability of such attacks, however, hinges on the feasibility of achieving the desired bit flips, a central challenge to the effectiveness of LeapFrog gadgets in real-world scenarios. In this scenario, flipping the PC from \texttt{0x555555555478} to \texttt{0x555555555578} only requires a 1-bit flip, which is a reasonable assumption for Rowhammer.
\begin{comment}
\begin{table*}[!htbp]
\caption{Comparison of Original and Alternative C Code with Corresponding Assembly Code}
\label{table:code_comparison}
\centering
\begin{tabular}{|p{0.45\linewidth}|p{0.45\linewidth}|}
\hline
\textbf{Original Code} & \textbf{Alternative Code} \\
\hline
% Original C code and Assembly
\begin{lstlisting}[language=C, caption={Original C code example}, label=lst:c_sample]
unsigned char message[32] = "This is a message to be signed";
int ret = send(client_fd, message, sizeof(message), 0);
\end{lstlisting}
\begin{lstlisting}[language={[x86masm]Assembler}, caption={Assembly for original C code}, label=lst:asm_sample]
0x0000555555555413: movdqa xmm0, [rip+0xce5]
0x000055555555541b: mov edx, 0x20
\end{lstlisting}
&
% Alternative C code and Assembly
\begin{lstlisting}[language=C, caption={Alternative C code example}, label=lst:c_sample_2]
unsigned char *message = "This is a message to be signed";
int ret = send(client_fd, message, strlen(message), 0);
\end{lstlisting}
\begin{lstlisting}[language={[x86masm]Assembler}, caption={Assembly for alternative C code}, label=lst:asm_sample_2]
0x0000555555555413: lea r14, [rip+0xc6e]
0x000055555555541a: lea r13, [rsp+0x60]
0x000055555555541f: mov rsi, r14
\end{lstlisting}
\\ \hline
\end{tabular}
\end{table*}

\end{comment}

However, tiny variations in the C code can change the resulting assembly code significantly. For example, consider the first approach for a TLS handshake, where the process allocates memory for a message to be signed. The C code and its corresponding assembly code are shown in Listing \ref{lst:combined_sample_1}. Alternatively, using a different method to allocate memory for the message results in a variation in the assembly code. This alternative approach and its corresponding assembly code are presented in Listing \ref{lst:combined_sample_2}.

\begin{figure}[!t]
\begin{lstlisting}[frame=single,
language=C++,
caption={Combined C and Assembly code for original memory allocation},
label={lst:combined_sample_1}]
// C Code
unsigned char message[32] = "This is a message to be signed";
int ret = send(client_fd, message, sizeof(message),0);

// Assembly Code
0x555555555413: movdqa 0xce5(%rip),%xmm0
0x55555555541b: mov $0x20,%edx
\end{lstlisting}
\end{figure}

\begin{figure}
\begin{lstlisting}[frame=single,
language=C++,
caption={Combined C and Assembly code for alternative memory allocation},
label={lst:combined_sample_2}]
// C Code
unsigned char *message;
message = "This is a message to be signed";
int ret = send(client_fd, message, sizeof(message),0);

// Assembly Code
0x555555555413: lea 0xc6e(%rip),%r14
0x55555555541a: lea 0x60(%rsp),%r13
0x55555555541f: mov %r14,%rsi
\end{lstlisting}
\end{figure}

\begin{figure}
    \centering
    \includegraphics[width=0.6\columnwidth]{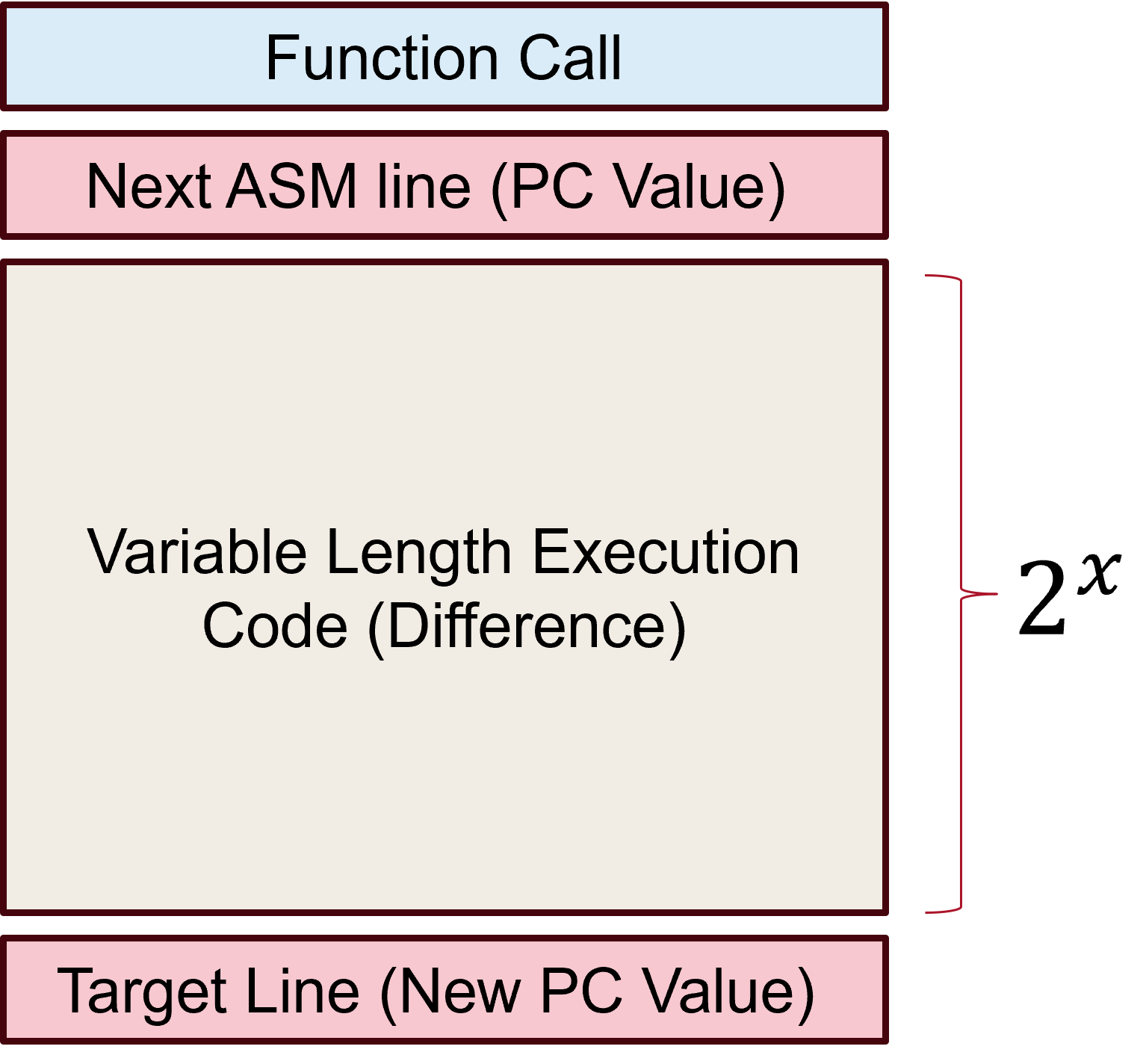}
    \caption{The best LeapFrog gadgets require a single-bit flip, where the distance between the two lines of code is a power of 2.}
    \label{fig:jumpdef}
\end{figure}

In the source code space, the alternative approach (Listing \ref{lst:combined_sample_2}) takes \texttt{0x55555555541f} - \texttt{0x555555555413} or 12 bytes of instructions, while the original approach (Listing \ref{lst:combined_sample_1}) occupies 8 bytes of assembly instruction (this excludes the size of the last instruction). Given the assumption that only one bit per page can be reliably flipped, identifying useful instruction skips that require a single bit change, as illustrated in Figure \ref{fig:jumpdef}, is crucial. This example illustrates the challenge of manually inspecting source code to determine the impact of tiny variations on assembly instruction distances. Hence, profiling binaries becomes an important tool in this context.

% TODO: Add example of how tiny C code change results in 4-1 bit flip [DONE]

\smallskip
\noindent
\subsection{Offline Memory Profiling}
%suitable for Rowhammer attacks

\begin{figure}[h]
    \centering
    \includegraphics[width=\columnwidth]{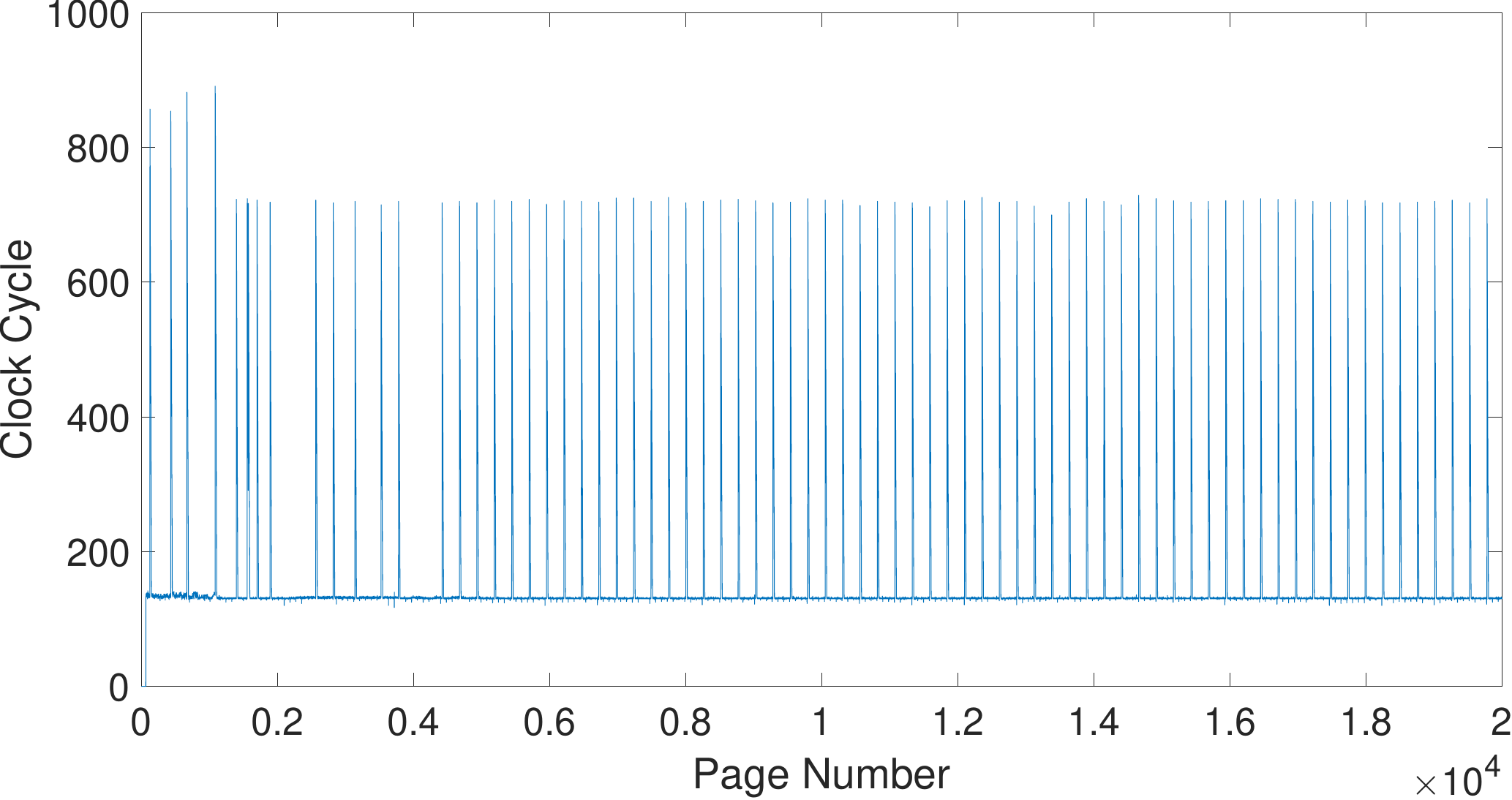}
    \caption{Timing peaks on virtual addresses detected by SPOILER~\cite{islam2019spoiler} attack. Virtual addresses on the peaks are contiguous on physical address space.}
    \label{fig:spoiler_peaks}
\end{figure}

\smallskip
\noindent
{\bf Finding Contiguous Memory} Virtual to physical address mappings are stored in \textit{pagemap} file in Linux OSs and it requires root privileges to access these translations. Using the SPOILER tool~\cite{islam2019spoiler}, we can reliably leak the information about the first 8 bits in physical addresses after the page offset bits. This allows us to find contiguous memory chunks in physical address space. In Figure~\ref{fig:spoiler_peaks}, the page numbers with the peaks in the y-axis are contiguous pages' physical address space. 

 \smallskip
\noindent
{\bf Finding Memory In the Same Banks} To find physical memory pages within the same bank, we employ techniques first described in \cite{pessl2016drama}. DRAM is structured in multiple banks that are physically isolated from each other, so while SPOILER can give an attacker physically contiguous chunks, the memory is distributed across multiple banks. This is a problem because hammering rows that are not adjacent within the same bank will not result in bit flips. 

We use the row conflict side channel to find co-located memory address accesses within the same bank. We take measurements by iterative reading between two addresses, and if the reading results in high latency (assuming we are not hitting cache), it means DRAM is clearing the row buffer and the addresses are located in the same bank. 

Alternatively, if we repeatedly read from two addresses and they are in different banks, the values will be loaded into their respective row buffers for their respective banks, and the reading time will have a lower latency.

\smallskip
\noindent
{\bf Executing the Rowhammer Bit Flip in a Many-Sided Context} Despite modern mitigation techniques against Rowhammer like Target Row Refresh (TRR), we are still able to induce flips in DDR4 memory by using a many-sided \cite{frigo2020trrespass} approach.
%\paragraph{Mechanics of Bit Flipping in a Many-sided Attack}

In the final phase of our attack, the task is to induce bit flips in the target memory location. This step marks the culmination of the profiling and memory manipulation processes. The challenge lies in the fact that while we can ascertain the occurrence of bit flips in a given row (a row that we deem ``flippy"), pinpointing the exact memory bits affected after the attack is not straightforward. This is due to the inherent nature of Rowhammer, where the attacker does not possess direct control over the specific memory areas being altered.

% TODO: EXPAND on this in tools section (process output)
However, the success of the attack is often evident through observable changes in the process' state. For instance, a successful execution might manifest as an unauthorized bypass of security measures, or broken encryption output. This indirect outcome serves as a confirmation of the attack's effectiveness. We further expand on this in section \ref{sec:autodetect}.

% TODO: list all practical challenges for LeapFrog attack. it should connect this section to the following sections (Locating PC).

% Talk about how minor changes in code allow for single bit flips...

\section{Locating the PC in the Stack}
% For rowhammer we need memory massaging...

% \smallskip
% \noindent
% {\bf Bait Page Analysis for Program Counter Targeting}%The initial approach to bait page profiling was designed for stack and register attacks under less dynamic conditions, preceding the complexities introduced by ASLR. This technique's primary goal was to identify and release bait pages—memory pages filled with irrelevant data—to strategically position the target variable within a flippable bit's reach in a victim's process memory.

To flip bits in the PC value with Address Space Layout Randomization (ASLR) enabled, the page that contains the PC value needs to be placed into the page with the bit that will flip during the Rowhammer attack. To do this, we use a method similar to that proposed in \cite{flipfengshui} where we deallocate a series of pages from the attacker process, launch the victim process, and experimentally determine some probability that the target data (in this case the PC) lands in the target location (the row with the flippy bits). We term the deallocation of pages ``baiting" in this paper. 

The profiling to determine the proper number of bait pages starts by allocating pages within the attacker's process space, designated to be released as bait. The procedure involves releasing a substantial number of bait pages, recording their physical addresses, and then correlating these with the physical address of the target variable in the victim process. The number of pages consumed by the victim process before allocating the target variable was determined through this correlation.

In a recent work~\cite{adiletta2023mayhem}, the victim's source code was altered to assign a unique value to the target register or stack variable, thereby making it identifiable in the memory during the profiling stage. This method is not possible with PC values as they are dependent on the compiler, so we introduce a new method to determine the number of bait pages required for the PC value. 

The dynamic nature of the PC under ASLR implemented in the Linux kernel necessitates a novel approach that involves identifying invariant values within the stack that serve as reliable \textit{fingerprints}. These fingerprints are used to determine the PC's offset relative to these constants, thereby facilitating the estimation of the required number of bait pages for effective targeting.

% TODO: Histogram of tally file 

\begin{figure}
    \centering
\includegraphics[width=0.6\columnwidth]{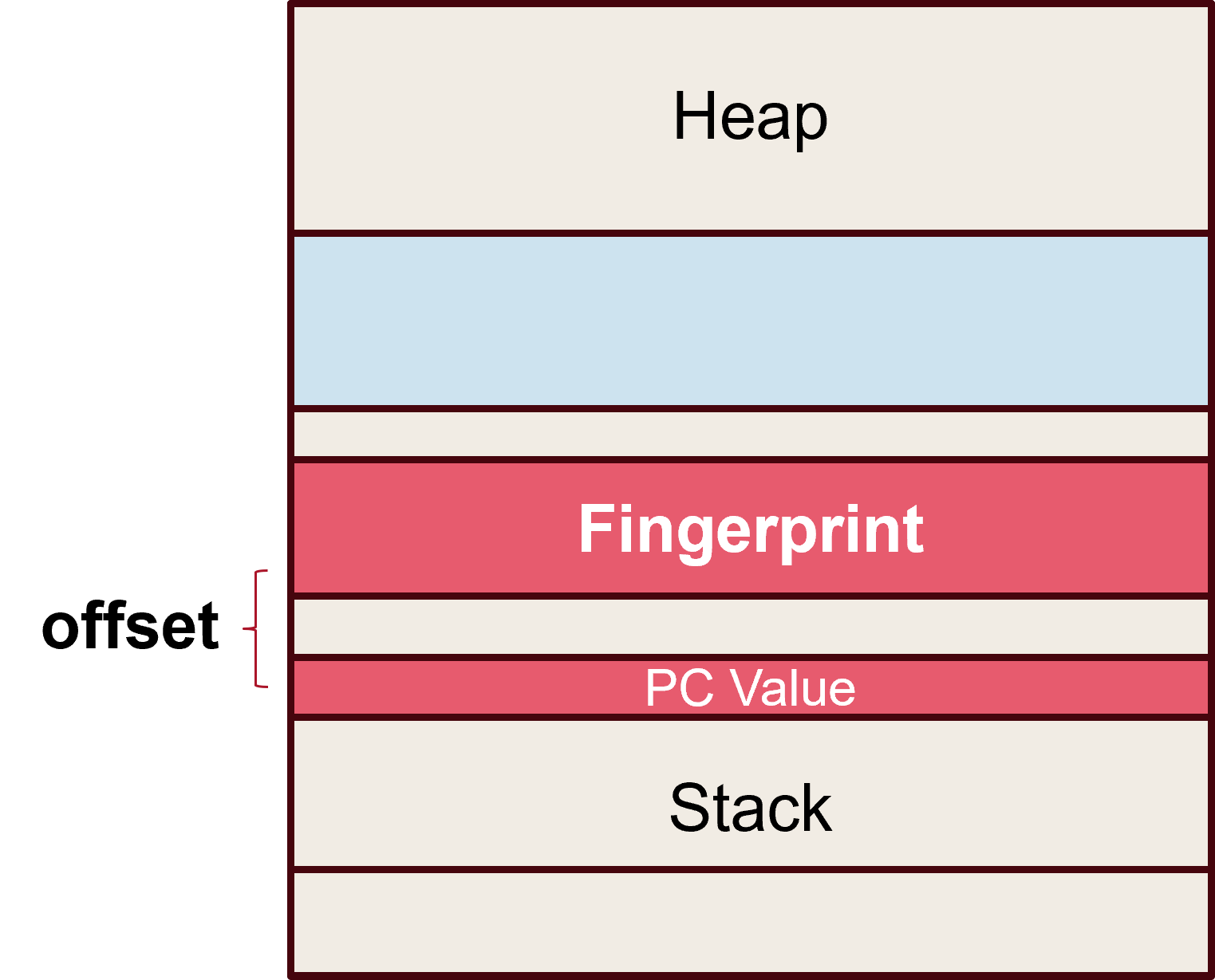}
    \caption{Once the fingerprint is located, there is a constant offset from the fingerprint regardless of ASLR, and this can be used for bait page profiling for the eventual attack}
    \label{fig:fingerprint2}
\end{figure}

\begin{figure}
    \centering
    \includegraphics[width=\columnwidth]{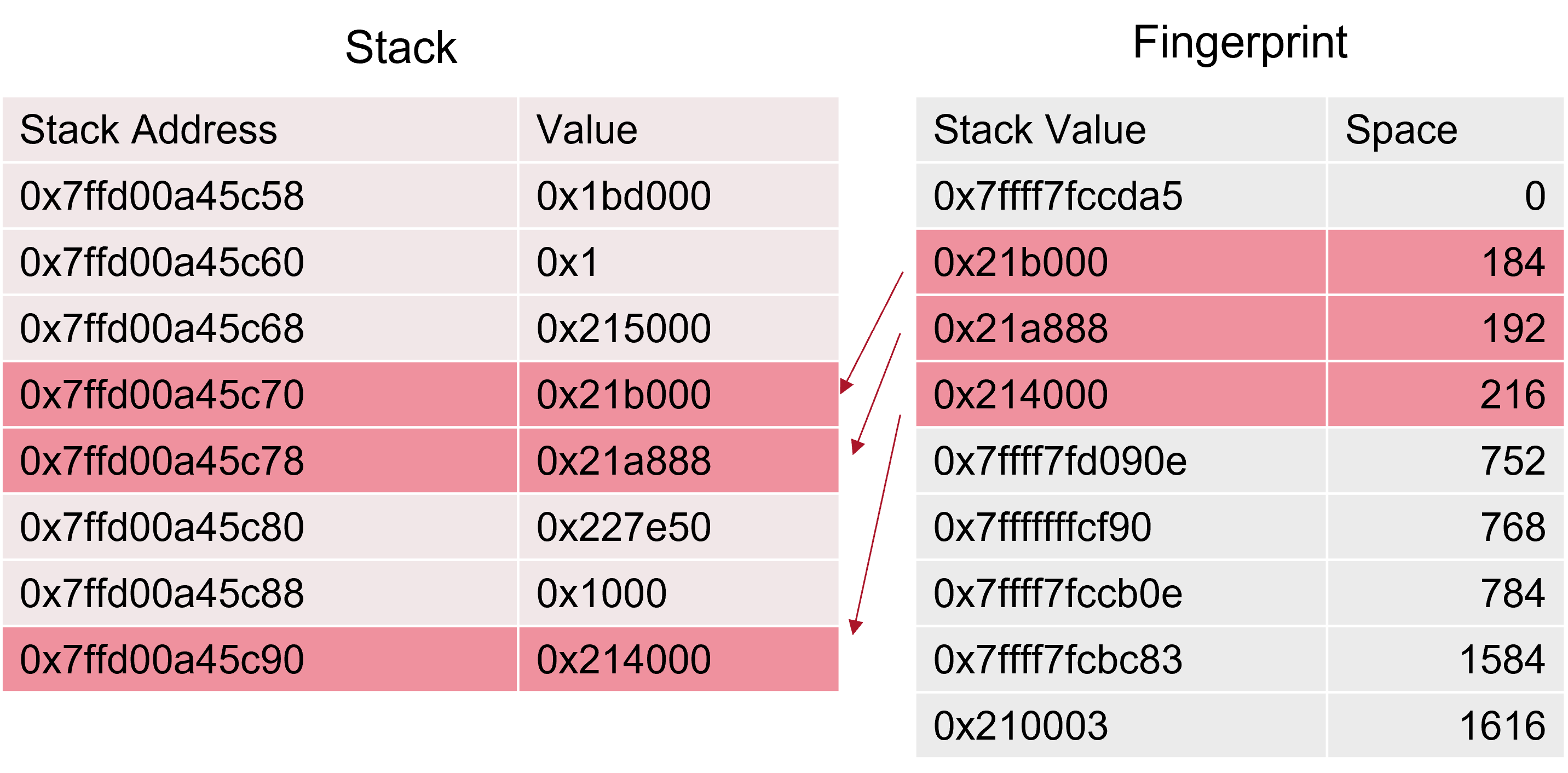}
    \caption{Finding constant values in the stack to create a fingerprint}
    \label{fig:fingerprint}
\end{figure}

\subsection{Fingerprinting the Stack}
% TODO: Give a fingerprinting example...

As the PC's address and value fluctuate with each process execution due to ASLR, our strategy leverages the relative stability of certain stack values and correlates an offset from those values. We first profile with ASLR disabled, knowing the target PC value in the stack from an assembly dump with GDB. We then determine an offset from the fingerprint as seen in Figure \ref{fig:fingerprint2}. Then with ASLR enabled, even with the PC value changing, the fingerprint remains identifiable and the offset from the fingerprint remains constant. The outcome is a refined understanding of the number of bait pages required to strategically position the PC, thus enhancing the precision of our Rowhammer attack in an ASLR-enabled environment. Fingerprinting only needs to be done once and is machine-independent. It is part of the offline stage of the attack.

The process begins by capturing snapshots of the stack at different instances and identifying unique values that persist across these snapshots. We implemented a Python script to automate this analysis. The script compares consecutive stack states, isolating values that remain unchanged— these become features of our fingerprints as seen in Figure \ref{fig:fingerprint}. By calculating the address differences between these consistent values and tracking their occurrence across multiple iterations, we build a comprehensive profile of the stack's layout. This profile is instrumental in pinpointing the location of the PC relative to the identified fingerprints and is versatile enough to be used on virtually any binary.

% TODO: Think about it later...

% \subsection{Examples...}
% Run the tool on... (Include table/visuals)

% Include C code where we are jumping...

% - TLS handshake PoC
% - OpenSSL
% - sudo

%%%%%%%%%%%%%%%%%%%%%%%%%%%%%%%%%%%%%%%%%%%%%%%%%%%%%%%
\section{Automatic Detection of LeapFrog Gadgets with \tool}\label{sec:autodetect}
%%%%%%%%%%%%%%%%%%%%%%%%%%%%%%%%%%%%%%%%%%%%%%%%%%%%%%%

\begin{figure*}
    \centering
    \includegraphics[width=\textwidth]{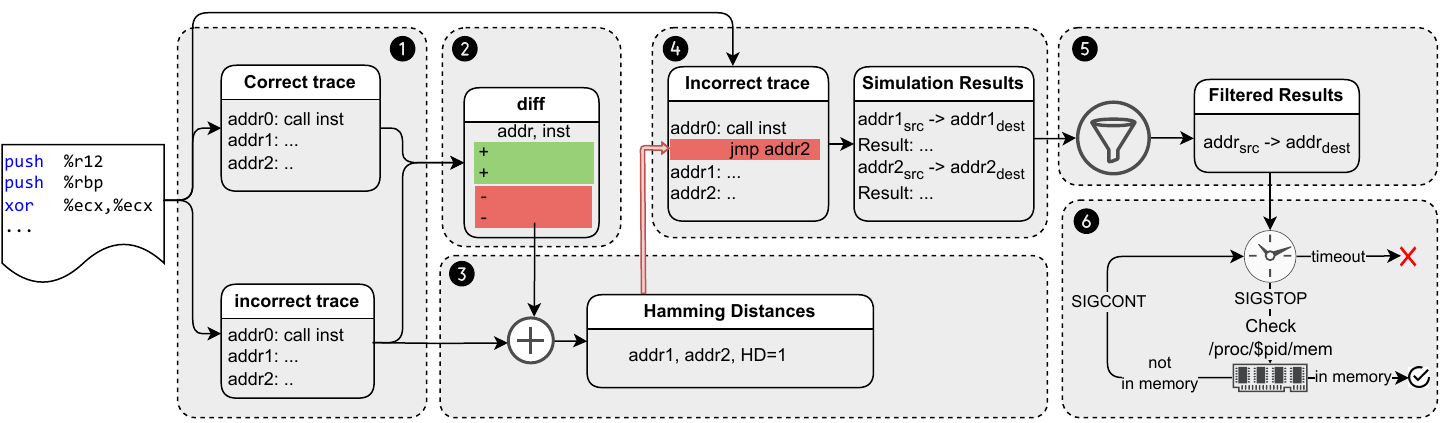}
    \caption{LeapFrog gadget detection using \tool framework}
    \label{fig:mfs}
\end{figure*}

Based on how the LeapFrog gadgets occur in the binary described in section \ref{sec:LeapFrog_attack}, we develop a custom tool we call \tool (Multidimensional Fault Simulator) that relies on dynamic binary instrumentation and analysis. Since the attack happens on program counters and registers, which are invisible to high-level code, such as  C/C++, it is not possible to do a static analysis of the source code.
We put together a set of rules that enables us to collect, filter, and pinpoint the potential LeapFrog gadgets. The overall design is shown in Figure~\ref{fig:mfs}.
% \begin{itemize}[leftmargin=*]

\blackcircled{1} First, \tool collects the instruction traces, specifically, the address of instructions executed, for different inputs. To detect the gadgets that cause security exploits, \tool chooses critical input pairs that cause differences in the program's control flow. Such inputs can be correct/incorrect private key pairs or passphrases for authentication programs. Together with the instruction addresses, we collect the execution time of each function executed. Since the return addresses of the functions with larger execution times will stay in the memory for a longer duration, they are potentially more viable targets.

\blackcircled{2} \tool then computes the difference between two instruction traces to find the instruction addresses that are executed with correct input(s) but not executed with incorrect input(s). Note that this is an optional step to reduce the complexity of the following steps, and it comes with a cost of false negatives. Moreover, depending on the program and type of exploit, it may not always be possible to get multiple different traces; see section \ref{sec:openssl_attack}. Alternatively, the whole instruction trace can be considered instead of only the difference.

\blackcircled{3} Regarding the choice of the fault model on PC values, the probability analysis in previous work~\cite{tol2023dont} is relevant. Given a sequence of bit offsets ${b_0,b_1,...,b_{k+l-1}}$ within a memory page and assuming a defective memory cell can flip solely in one direction, the conditional probability of identifying a compatible target page $t$ amongst $N$ susceptible pages can be expressed as: \begin{multline}\label{eq:prob}
    p\big(t|\{b_{n_{0\rightarrow1}}\}\in\{0\rightarrow1\},\{b_{n_{1\rightarrow0}}\}\in\{1\rightarrow0\}\big) =\\
    1- \bigg(1- \prod_{i=0}^{k-1}{\dfrac{n_{0\rightarrow1}-i}{S-i}}\times\prod_{j=0}^{l-1}{\dfrac{n_{1\rightarrow0}-j}{S-k-j}}\bigg)^N,
\end{multline}

where $n_{0\rightarrow1}$ and $n_{1\rightarrow0}$ represent the average counts of error-prone cells on a page that can be flipped from $0$ to $1$ and from $1$ to $0$ respectively, $k$ and $l$ denote the counts of bit locations needing flips from $0$ to $1$ and $1$ to $0$ respectively, and ``$S$" signifies the total bit count per page. 
% We further analyze Equation~\ref{eq:prob_reduced} with the numbers calculated in Table~\ref{tab:flip_profiles}. 

\begin{figure}[h]
    \centering
    \includegraphics[width=\columnwidth]{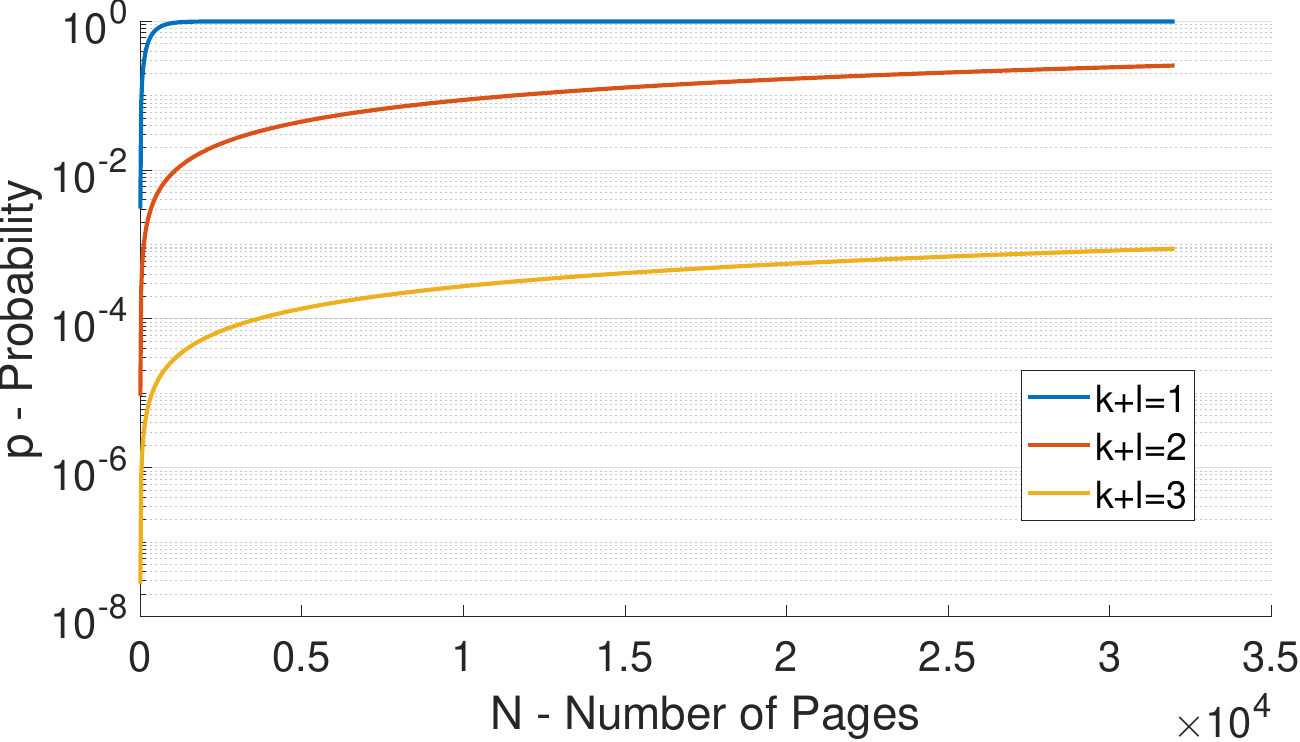}
    \caption{Probability of finding a page among N pages for different k+l values. k+l states the number of targeted bit offsets in a page.}
    \label{fig:prob1}
\end{figure}

We evaluate the likelihood of locating a specific target page $t$ for various values of $N$ and three distinct configurations of $k+l$, where $k+l$ represents the number of bit offsets within a single page. As shown in Figure~\ref{fig:prob1}, when there is 1 bit per page, approximately 2200 pages are sufficient to reach 99.99\% accuracy in a DDR4 system with an average of 100 bit flips per page. For configurations with 2 and 3 bits per page, using the same number of pages results in probabilities of 2\% and 0.006\%, respectively.

\tool looks for address pairs that hold the following conditions:

\begin{equation}\label{eq:Hamming}
    d_{HD}(\textrm{addr}_{\text{exec}}^i, \textsf{addr}_{\text{return}}^j) = 1
\end{equation}

where $addr_{\text{exec}}^i$ is the address of the $i^{th}$ instruction that is executed, $addr_{\text{return}}^j$ is the return addresses of the $j^{th}$ call instruction, and   $d_{HD}$ is the Hamming distance between two addresses. $i$ and $j$ are bounded by the number of all instructions executed ($n$) and the number of call instructions executed ($m$), respectively. Although this operation has $O(m^n)$ complexity, it can be implemented with bitwise xor and can be parallelized using multiple processor cores. The condition given in Equation~\eqref{eq:Hamming} is determined by the Rowhammer fault model. Given that finding a suitable memory page in memory is only realistic with single-bit flip fault models, \tool assumes we can only flip a single bit. Yet, the method is generic enough to cover other potential fault models, such as optical fault injection or electromagnetic fault injection, where multiple-bit flips are more likely~\cite{breier2022practical}. This step generates a list of pairs of addresses in the following format: $\{<addr_{src}^k, addr_{dest}^k>\}$ where $addr_{src}^k$ is the $k^{th}$ instruction address that \tool targets in the binary's execution with the input that we want to affect the control flow of, such as an incorrect private key, and $addr_{dest}^k$ is the corrupted instruction address after fault injection.

\blackcircled{4} For each address pair we get from the list generated in the previous step, \tool starts a simulation session. \tool executes the binary again with the incorrect input and simulates a bit flip on the instruction address $addr_{src}$ to make it $addr_{dest}$. Certain instructions may be executed multiple times in a single execution. To correctly cover that case in our fault model, we keep a counter variable for a specific instruction that increments every time the binary executes the same instruction. In a single execution of the original binary, if an instruction is executed $N$ times, we attempt the fault simulation $N+1$ times, until we no longer see the same instruction in the trace. 

\blackcircled{5} After the bit flip simulation, \tool continues the execution of the binary without further faults and observes the new behavior. The analysis of the new behavior is not a trivial task. There are several options where we can observe changes compared to the original execution. For instance, we can observe changes in the total number of executed instructions, the number of instructions that match with the correct input execution trace, the return code of the program, outputs to standard streams, ports that are accessed, function calls, authentication results, etc. The choice of observable depends on the program under test. In this work, \tool uses the return codes, standard outputs/errors, and authentications on different case studies.

% \end{itemize}

% \smallskip
% \noindent
% {\bf Linux Kernel and GDB}

\blackcircled{6} Once \tool has a list of PC values that potentially result in misauthentication or bypass with a single-bit flip, it then evaluates if they are practical to attack from a timing perspective. In some cases, a single bit flip will result in the desired behavior in a process but the attack window of time is too short to effectively attack the target. Additionally, the attack window needs to be long enough to allow for noise in the system - as processes will often take a variable amount of time to execute and get to the vulnerable area in the code where the PC value is shelved in the stack. 
\tool uses process degradation to increase the viability of LeapFrog gadgets, as slowing down a process artificially increases the attack window time. % But in general, most LeapFrog gadgets that are found in a process will not be practical to attack because they either require too many bit flips or the attack window is too short.
Note that this step is system-specific and it can be affected by the current processor/memory load. Although it is necessary to find viable targets in the list for an end-to-end attack, it does not guarantee that the other targets are not viable in different system configurations, or different systems.

\tool starts the victim process and then immediately stops it with a SIGSTOP signal and it checks if the PC value is currently in the stack of the process. %It is fair to assume that the target processes can be stopped with SIGSTOP and started again with SIGCONT because our targets are owned by the non-root attacking user.
If not, the process is killed and restarted, and stopped after a slightly longer period, in a process we call time sweeping. The challenge is sending a SIGSTOP with the highest timing precision possible. Different implementations of signals will yield different timing resolutions. For example, Python has a signal library that can be used to generate signals similar to a bash script, but there is considerable delay and imprecision in the time it takes to send a signal. 

\subsection{Tool Implementation}

We used Intel's dynamic binary instrumentation framework, \texttt{Pin} \cite{luk2005pin}, which allows for process analysis without altering its core behavior to implement \blackcircled{1} and \blackcircled{4} of \tool. Using Pin also makes it possible to find LeapFrog gadgets in binaries that do not have a source code since it does not require recompiling.
%The integration with Intel Pin significantly enhances this tool's capabilities by utilizing the dynamic analysis strengths of Pin. %Pintool is designed for detailed binary instrumentation, enabling researchers and developers to analyze program execution at a granular level. 
In the context of \tool, Pin's capabilities are harnessed to monitor the execution trace of a binary. This integration allows for a thorough analysis of potential LeapFrog gadgets by observing how changes in PC values influence program behavior. For each executed instruction, our tool outputs the virtual address of the instruction and disassembly of the machine code. If the instruction is a call instruction, it also outputs the return address of the call. The return address of the call is usually the PC value that is pushed onto the stack before executing the called routine. For every write to STDOUT and STDERR, the tool forwards a copy of the buffer to a text file for further analysis. To avoid the effect of overhead caused by instruction-based instrumentation, function timings are collected in a separate session on every function entry and exit.

\blackcircled{2} is a simple comparison operation on the correct and incorrect execution traces implemented with \texttt{diff} command line tool in Linux. 

\blackcircled{3} is implemented in Python. \tool parses the instruction traces and computes the Hamming distance between the return addresses and instruction addresses of all executed instructions in the correct trace or the list of addresses we get from \blackcircled{2}. The Hamming distances are calculated using the native \texttt{bit\_count} function in Python followed by \texttt{bitwise\_xor} in \texttt{numpy} library. The operation is parallelized on multiple cores to speed up the analysis.

The bit flip simulation part of \tool (\blackcircled{4}) is done using \texttt{Pin} which takes the address pairs and simulates every fault independently. The faults on PC values are implemented as direct jumps to the corrupted addresses by adding \texttt{jmp $addr_{dest}$} after function returns.
Since we add a direct jump to the target address by injecting a line of assembly with the Pin tool, it is functionally equivalent to corrupting the PC value in memory.

\blackcircled{5} filters the simulation results depending on the program and targeted exploit type. For different types of exploits, we filter by return code (section \ref{sec:tls_attack}), value in STDOUT (section \ref{sec:openssl_attack}).
%For the time profiling section of the automated gadget detection tool, the list of PC values that only require a single bit flip is checked one at a time through a time-sweeping operation. 

% TODO: Rewrite this section to fit better with last 6 section [DONE]

\blackcircled{6} The last stage of \tool takes the list of PC values generated from the previous steps and determines which are practical from a timing perspective. It does this by sweeping the process in the time domain determining when it needs to stop the process to find particular PC values in the stack. 

We begin by defining when we want to start our sweep, and what interval we want to sweep at. For OpenSSL as an example, we started our sweep at 0ns and had an interval of 100ns. Generally, the higher the resolution of the sweep, the longer the simulation takes. However, a smaller interval increases the likelihood that we will successfully send a SIGSTOP at a time when the target PC value is in the stack. 

To determine if a PC value is in the stack, we start the victim process as a non-root user on a sibling core to a core that we are attacking with SMC to degrade the performance. For example, in our tests we isolated cores 6 and 14 and triggered SMC events on core 14 while running the victim process on core 6. Once we have the process identification number (PID) of the process and send a SIGSTOP, we use the Linux process interface to check the stack for the PC value. We do this by looking at the \texttt{/proc/[pid]/maps} file to determine which offsets in the victim process's address space contain the stack, and then we read from \texttt{/proc/[pid]/mem} at the offsets determined by \texttt{/proc/[pid]/maps} to find the PC values. The tool will generate a dictionary of stack addresses/values for the victim process that we can search through. 

If during a sweep the tool finds the PC value in the stack, it will simulate a flip by overwriting that value with the new PC value determined by the previous steps to verify that the gadget does result in the intended behavior (privilege escalation, data leak, mis-authentication, etc...). 

Generally, if \tool can successfully pass all stages of filtering with a particular LeapFrog gadget, we believe that it can be attacked and flipped with Rowhammer to cause the desired behavior.

\section{Experiments}
%%%%%%%%%%%%%%%%%%%%%%%%%%%%%%%%%%%%%%%%%%%%%%%%%%%%%%%
\smallskip
\noindent
{\bf Experiment Setup}
The experiments are conducted on a system with Ubuntu 22.04.2 LTS with 6.2.0-37-generic Linux kernel installed. The system uses an Intel Core i9-9900K CPU with a Coffee Lake microarchitecture. We used a dynamic clock frequency instead of a static clock frequency to improve the practicality of the attack.
End-to-end attack experiments are performed on a single DIMM Corsair DDR4 DRAM chip with part number CMU64GX4M4C3200C16 and 16GB capacity and TRR enabled. DRAM row refresh period is kept at 64ms, which is the default value in most systems.
In all the experiments, we used 100s simulation timeout, since the fault simulations rarely cause infinite loops.
We empirically observe that using the Python signals library, the target process could complete 34M cycles before the attacker can stop it, with a standard deviation of 2.7M cycles. Alternatively, using a bash script, the victim process can only complete 18M cycles before it is stopped, with a standard deviation of 0.3M cycles. There is an order-of-magnitude difference in precision stopping a process with bash vs. with Python.

{\bf Overview of \tool Results}
For ML Misclassification, we generated Table 1 which shows ~1300 potential candidates 1 hammering distance away, and of those 23 result in misclassification. For OpenSSL, we found 36 ciphers that are exploitable via Leapfrog to leak plaintext. For PQC, we list the verification bypass gadgets in the last column, with a total number of gadgets that result in mis-authentication between all the signature schemes being 177 gadgets."

{\bf Note about Candidates}
Our tool finds gadgets that skip crucial instructions. We then simulate each skip to check if it leads to code that leaks plaintext, accepts invalid signatures, or bypasses checks. Not every candidate is a fully exploitable given timing constraints.  In our TLS example, ~2,500 potential one-bit flips produce about 20 confirmed gadget paths that bypass signature validation.

% Describe attacking these programs:

% Include C code where we are jumping...

% - TLS handshake PoC
% - OpenSSL
% - sudo

% Heatmap of flips in kernel

%/home/berksunar/stack_attack/Dropbox/LeapFrog/tls_handshake_attack/logs/rowhammer_log_2023-11-29_00-24-19.txt
%Found target data at v_addr 7ffed62987b8
%Physical address of target data: 185f70000
%Flippy physical address: 185f70000
%Target physical address: 185f70000
%Flippy Success: 55d1bb00c478 v_addr:7ffed62987b8 %phys_addr:185f70000
%Old PC value: 55d1bb00c478
%New PC value: 55d1bb00c578
%PC value changed! ##############

% important TODO: Results from experiment <- ANDREW
% Include paragraph on baiting

% Include table of results...
% # of flips
% offline/online time
% baiting success
% Number of PC value flips

%For example, the \texttt{sudo} process is owned by a non-root process and can be stopped with SIGSTOP without root access, and attacked to gain root access. The OpenSSL processes are also owned by the attacking process and can then be stopped with SIGSTOP.

\subsection{ML Misclassification}
In this section, we investigate the potential implications of instruction skipping in the machine learning domain, specifically for decision tree algorithms.
A decision tree is an ML model used to make predictions based on a series of binary choices, effectively splitting data into increasingly specific groups. It starts with a single node, which branches into possible outcomes based on the features of the data. Each branch represents a decision pathway, and each node in the pathway represents a test on a specific attribute. This process continues until a leaf node is reached, which provides the predicted outcome. They are widely used in various applications, from financial forecasting \cite{liu2023interpreting} to medical diagnosis \cite{shaik2023big} due to their interpretability, and efficiency for a variety of tasks such as classification, and feature importance ranking. We choose a decision tree for proof of concept yet instruction skipping attacks can be effective in every kind of model implementation.

Classification algorithms may be vulnerable to the LeapFrog attack under the threat model that an attacker is co-located on the server with the victim process running the model, and the attacker would like to force a particular output. If the attacker faults the victim process program counter and forces a jump in the code, the result may be a misclassification or a forced classification of a particular output. This attack is different from other Rowhammer attacks on machine learning models \cite{tol2023dont} because for this attack we do not need to know the model weights before hand, and we consider this a gray box model.

In this experiment, we use a public implementation\cite{decisiontree} as our target. We simulate program counter flips and observe the effects on the model output. We follow a similar procedure to previous examples, where we experiment with a hammering distance of 1, 2, and 3 and determine the number of successful LeapFrog gadgets with each of these distances. In Table \ref{tab:numberofgadgets_tree}, we can see various number of LeapFrog candidate gadgets that might result in a misclassification. After simulating these gadgets, we found 23 of the 1363 potential gadgets within 1 hammer distance would result in a misclassification.

\begin{table}[h]
    \centering
    \begin{tabular}{c|c |c| c| c| c}
    \toprule
        Target & Size & \#Inst.\textsubscript{exec}& $d_{HD}$ & \multicolumn{2}{c}{\# Candidates} \\
          &&& &  \blackcircled{2} on & \blackcircled{2} off\\

    \midrule
          \multirow{3}{*}{Decision Tree} &\multirow{3}{*}{99KB} &\multirow{3}{*}{38417} &  1 & N/A & 1363  \\
                        &&&  2 & N/A & 8667 \\
                        &&&  3 & N/A & 32326 \\

    \bottomrule
    \end{tabular}
    \caption{Number of gadget candidates found in decision tree algorithm with different Hamming distances.}
    \label{tab:numberofgadgets_tree}
\end{table}

\subsection{OpenSSL Encryption Bypass}\label{sec:openssl_attack}

We analyze \texttt{openssl} command line tool that uses OpenSSL v1.1.1w for block cipher and stream cipher implementations. 
For each cipher, we give a simple plaintext that contains the \texttt{helloworld} string and run encryption without salt with a simple passphrase. 
We aim to find LeapFrog gadgets in the binary that can be exploited for bypassing encryption steps in the ciphers, revealing the plaintext.

First, we scan the binary using \tool as described in section \ref{sec:autodetect}. Since we do not aim for any authentication bypass in this scenario, and the execution traces are deterministic for fixed inputs, step \blackcircled{2} is not applicable. Instead, in step \blackcircled{3}, we compare the return addresses in a single trace against all the instruction addresses in the same trace to look for targets with $d_{HD}=1$. This means that ultimately, for each return address, we test 12 different jumps, a trial for each bit in the page offset. Table \ref{tab:numberofgadgets_openssl} shows the total number of gadgets for each hammer distance, with a hammer distance of 1 having a total of 2700 candidates. 

We scanned the binary with 135 different ciphers available in OpenSSL. Most of the time the binary was not affected by the simulated bit flip and correctly produced the ciphertext.

\begin{table}[h]
    \centering
    \begin{tabular}{c|c |c| c| c| c}
    \toprule
        Target & Size & \#Inst.\textsubscript{exec}& $d_{HD}$ & \multicolumn{2}{c}{\# Candidates} \\
          &&& &  \blackcircled{2} on & \blackcircled{2} off\\

    \midrule
          \multirow{3}{*}{OpenSSL} &\multirow{3}{*}{818KB} &\multirow{3}{*}{49431} &  1 & N/A & 2700  \\
                        &&&  2 & N/A & 20208 \\
                        &&&  3 & N/A & 70475 \\

    \bottomrule
    \end{tabular}
    \caption{Number of gadget candidates found by \tool in for fault models with different Hamming distances. We ran OpenSSL with \texttt{aria-128-cbc} cipher.}
    \label{tab:numberofgadgets_openssl}
\end{table}

\begin{figure}
    \centering
    \includegraphics[width=\linewidth]{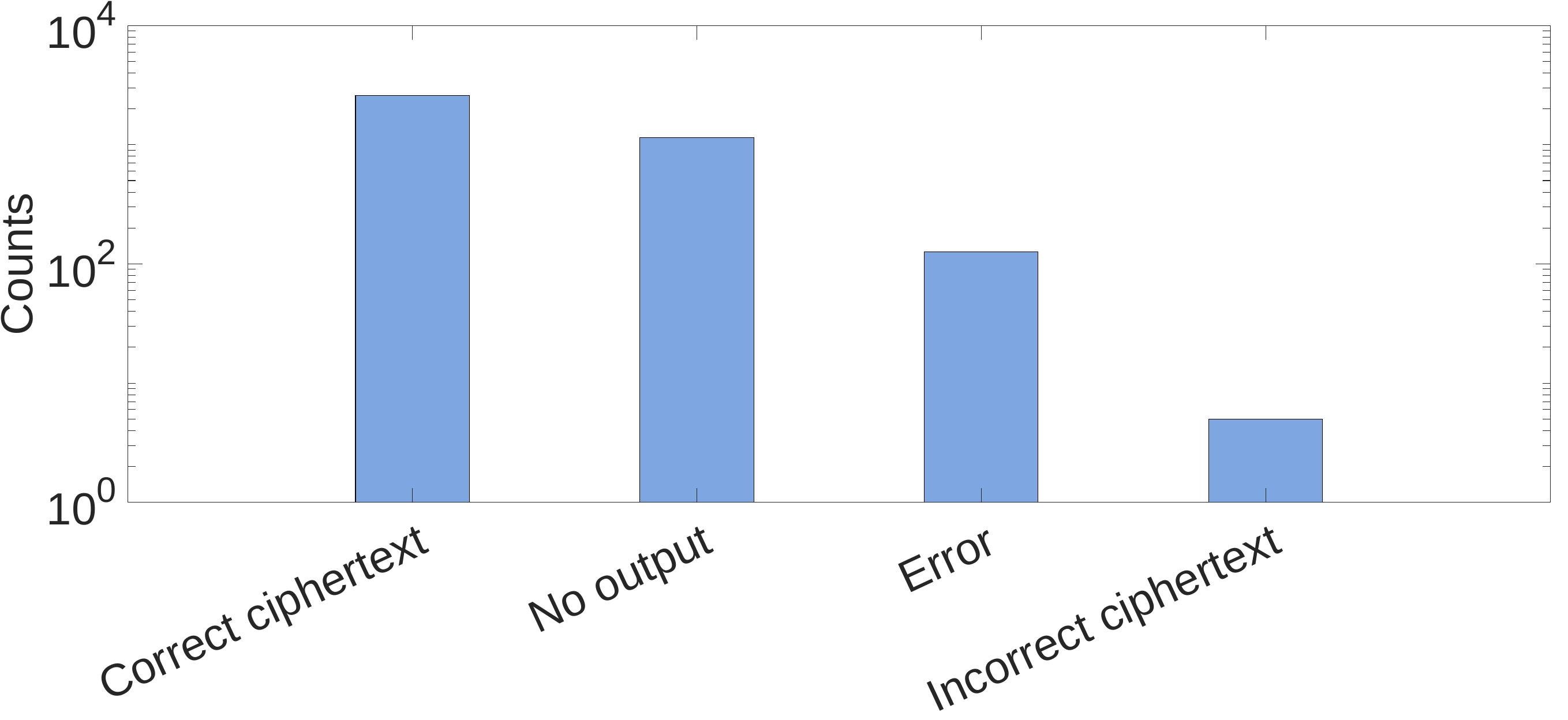}
    \caption{\texttt{aes-256-ctr} simulation results }
    \label{fig:aes256}
\end{figure}

\begin{figure}
    \centering
    \includegraphics[width=\linewidth]{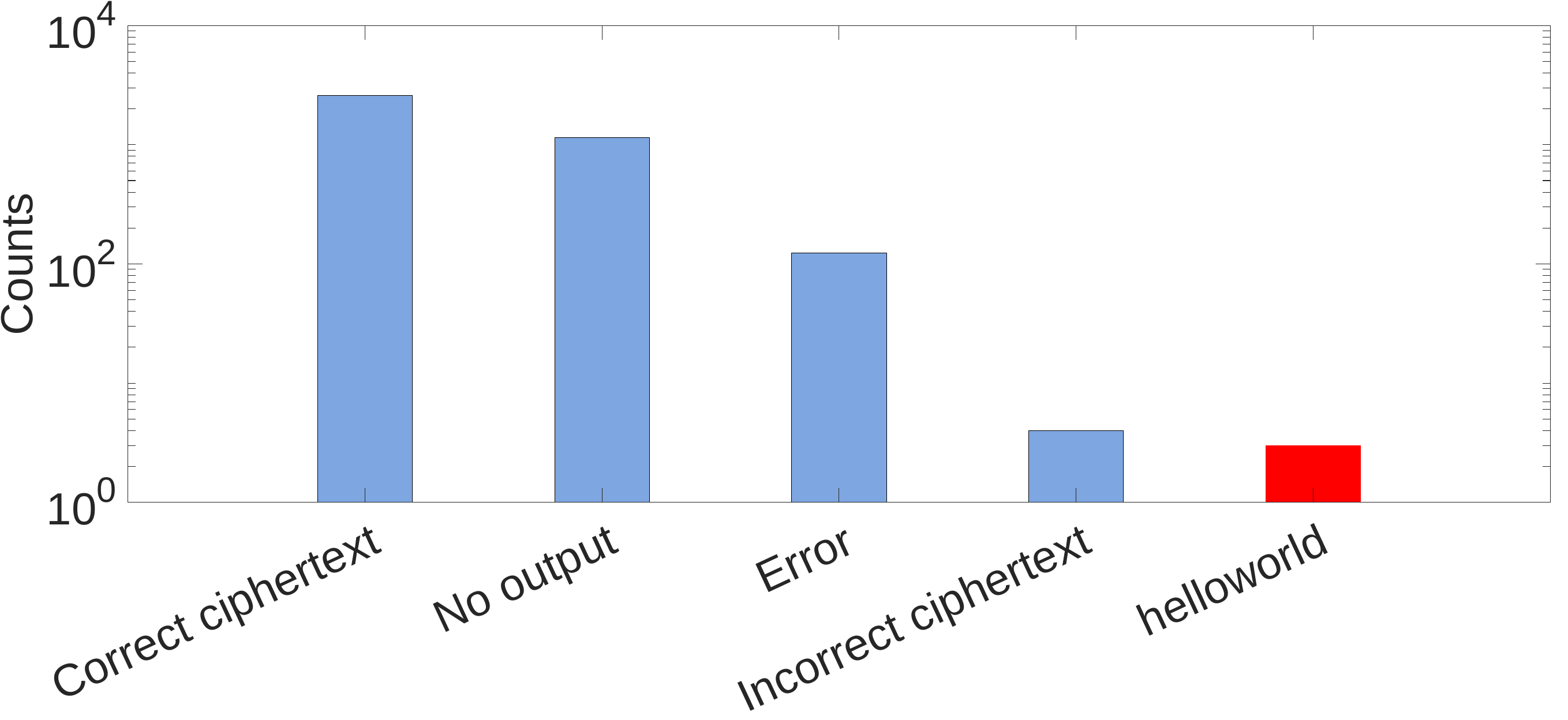}
    \caption{\texttt{aria-256-ctr} simulation results. Plaintext \texttt{helloworld} is revealed three times.}
    \label{fig:aria256}
\end{figure}

\begin{figure}
    \centering
    \includegraphics[width=\columnwidth]{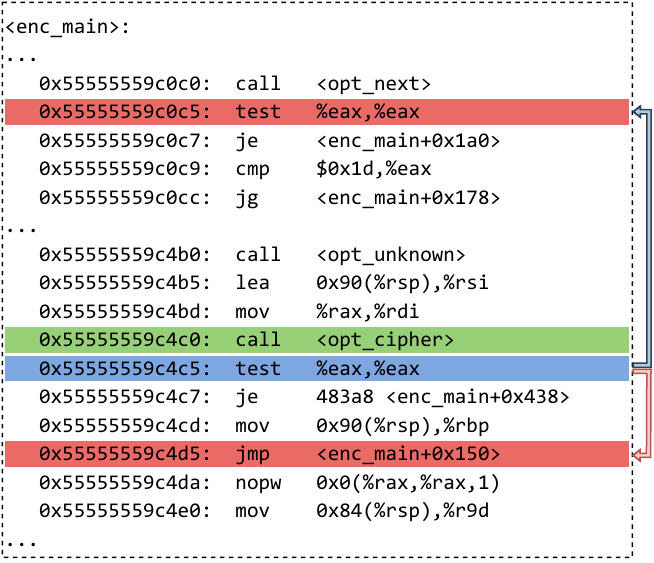}
    \caption{LeapFrog gadget in OpenSSL command line tool resulting in encryption bypass in aria-128-cbc block cipher. The PC value that fault is injected into, $addr_{src}$, is highlighted in \colorbox{blue}{blue}. The new value after the fault is injected, $addr_{dest}$, is highlighted in \colorbox{red}{red}. The fault is injected during the execution of the function call highlighted in \colorbox{green}{green}.}
    \label{fig:openssl_asm}
\end{figure}

Figure ~\ref{fig:openssl_asm} illustrates one of the LeapFrog gadgets found in the \texttt{openssl} command line tool.  When we corrupt a single bit in \texttt{0x55555559c4c5}, the return address of \texttt{opt\_cipher} function, to make it \texttt{0x55555559c0d5}, the function returns to the corrupted return address, skipping three instructions in between. Similarly, another single-bit corruption to (\texttt{0x55555559c0c5}) causes the function to return to an earlier point in the program. We verified that both of these bit flips cause the binary to skip the whole encryption and instead output the plaintext.  
Similarly, \tool detected LeapFrog gadgets that are used in 36 ciphers including block ciphers and stream ciphers. The ciphers with LeapFrog gadgets that revealed full or partial plaintext are listed in Table~\ref{tab:cipherslist}. 
Figure ~\ref{fig:aes256} and ~\ref{fig:aria256} summarize the simulation results for \texttt{aes-256-ctr} and \texttt{aria-256-ctr}  respectively.

Even with ASLR enabled, these gadgets are reproducible because ASLR does not randomize the last 12 bits of the code space (the page offset). We only simulated faults in the last 12 bits (which should be the same across all x86 machines the process is compiled for), thus, the LeapFrog gadgets should work across machines without the need for rescanning. 

\begin{table}
    \centering
    \footnotesize
    \begin{tabular}{c|c}

\toprule
    Recovered & Cipher \\
\midrule
 \texttt{\textcolor{darkpastelred}{helloworld}} & aria-128-cbc, aria-128-cfb,aria-128-cfb1\\
             & aria-128-cfb8, aria-128-ctr, aria-128-ofb \\
             & aria192, aria-192-cbc, aria-192-cfb \\
             & aria-192-cfb1, aria-192-cfb8, aria-192-ctr \\
             & aria-192-ofb, aria256, aria-256-cbc \\
             & aria-256-cfb, aria-256-cfb1, aria-256-cfb8 \\
             & aria-256-ctr, aria-256-ofb, bf-ofb \\
             & rc2-ofb, rc4, rc4-40 \\
 \midrule          
  \texttt{\textcolor{darkpastelred}{hellowor}}... & bf-cfb, rc2-cfb \\
 \midrule
 \texttt{\textcolor{darkpastelred}{hdlmowor}}... & idea-cfb, idea-ofb \\
 \midrule
 \texttt{\textcolor{darkpastelred}{oworhell}}... & bf, bf-cbc, bf-ecb, blowfish \\
  \midrule
\texttt{?\textcolor{darkpastelred}{rl}\#a?gy?...} & chacha20, des-ede3-ofb, des-ede-ofb, des-ofb \\
\bottomrule
    \end{tabular}
    \caption{36 ciphers implemented in OpenSSL that are vulnerable to LeapFrog attack. Each given cipher reveals the plaintext fully or partially in the ciphertext due to skipped encryption steps.}
    \label{tab:cipherslist}
\end{table}

\subsection{Post-Quantum Cryptography Schemes}\label{sec:pqc_attack}

\hl{NIST announced the standards for Post-Quantum Cryptography (PQC) in
%FIPS 203~\cite{raimondo2023module}, 
FIPS 204~\cite{raimondo2023signature}, and FIPS 205~\cite{raimondo2023hash}. These standards are used for 
%the key-encapsulation mechanism (FIPS 203) and 
digital signatures to protect against quantum attacks. We use Open Quantum Safe (\texttt{liboqs version 0.11.1-dev}) library~\cite{liboqs}, an open source library for PQC algorithms, to find \attack gadgets on FIPS standards using \tool tool.}

One of the algorithms selected by NIST for standardization is CRYSTALS-Dilithium, which serves as a digital signature scheme providing post-quantum security guarantees. Dilithium relies on the hardness of structured lattice problems, such as the Learning With Errors (LWE) problem, which is believed to be intractable for quantum computers. Another prominent algorithm is FALCON, which offers smaller key sizes and signatures by employing the NTRU lattice, making it a competitive choice for constrained environments. Our analysis of these algorithms reveals that, despite their robust design against quantum attacks, they still exhibit vulnerabilities at the implementation level, susceptible to hardware fault injections like the \attack used in Rowhammer-based exploits.

In digital signature schemes, we find gadgets that produce several failure modes in the Open Quantum Safe Library. The most critical error is a bypass of the signature verification. Note, that while we experimented with Post-Quantum encryption schemes, theoretically LeapFrog gadgets should work on classical encryption schemes as well. 

%One observed failure is the inability to create a new pthread, as indicated under the "Pthread Creation Fail" column. This issue occurs when a LeapFrog gadget disrupts the control flow during thread initialization, potentially leading to a denial-of-service (DoS) attack by halting the execution of cryptographic routines that rely on multithreading. As seen in Table \ref{tab:pqcdss}, all the signature schemes except MAYO-5 contain a gadget that results in the inability to create a PThread. 

The ``Magic Number Mismatch" column in Table \ref{tab:pqcdss} highlights instances where the injected fault corrupts memory regions containing predefined magic numbers used for integrity checks. This mismatch signifies unintended memory corruption caused by the LeapFrog gadget, which can lead to unpredictable behavior or system crashes. According to Table \ref{tab:pqcdss}, all signature schemes also contain gadgets for this failure mode, with Dilithium 3 containing the most number of LeapFrog gadgets. 

Failures during key generation (``Key Gen. Fail"), signature generation (``Sig. Gen. Fail"), and signature verification (``Sig. Verif. Fail") were also identified. Such failures can be exploited to disrupt normal cryptographic operations, resulting in denial-of-service (DoS) attacks or weakening cryptographic strength by producing invalid or insecure keys and signatures. All schemes contain this type of gadget. 

The ``Incorrect Verification" column denotes scenarios where invalid signatures are erroneously accepted as valid. This occurs when a LeapFrog gadget alters the control flow of the verification routine, enabling attackers to perform impersonation attacks by forging signatures that bypass standard validation checks.

\begin{figure}
    \centering
    \includegraphics[width=\columnwidth]{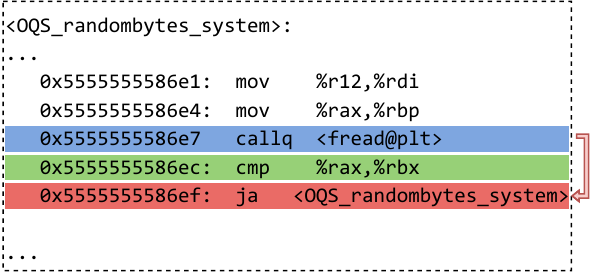}
    \caption{LeapFrog gadget detected in \texttt{liboqs} binary for Dilithium PQC Digital Signature Scheme. The PC value that fault is injected into, $addr_{src}$, is highlighted in \colorbox{blue}{blue}. The new value after the fault injected, $addr_{dest}$, is highlighted in \colorbox{red}{red}. The fault is injected during the execution of the function call highlighted in \colorbox{green}{green}.}
    \label{fig:dilithiumgadget}
\end{figure}

Lastly, the ``Verification Bypass" column in Table~\ref{tab:pqcdss} highlights instances where the signature verification routine can be entirely circumvented using LeapFrog gadgets. Similar to the TLS attack scenario described in Section~\ref{sec:tls_attack}, this allows an attacker to craft an invalid signature and have it accepted as valid at the client's end. By flipping bits in the Program Counter (PC) values using the \attack within the client's memory space, the attacker effectively bypasses the signature verification routine. This vulnerability poses a significant security risk by enabling impersonation attacks and facilitating unauthorized access or actions within the system. Notably, Dilithium3 exhibits the highest number of LeapFrog gadgets for this threat, indicating a greater susceptibility to such attacks. An example of such a LeapFrog gadget in Dilithium is seen in Figure \ref{fig:dilithiumgadget}.

\begin{table*}[htbp]
\centering
\small
\setlength{\tabcolsep}{2.5pt}
\begin{tabular}{lccccccccc}
\toprule
\textbf{Scheme} & \textbf{\# Instructions} & \bfseries\makecell{Candidate \\ Gadgets} & \bfseries\makecell{Magic Number \\ Mismatch} & \bfseries\makecell{Key Gen. \\ Fail} & \bfseries\makecell{Sig. Gen. \\ Fail} & \bfseries\makecell{Sig. Verif. \\ Fail} & \bfseries\makecell{Incorrect \\ Verification} & \bfseries\makecell{Verification \\ Bypass} \\
\midrule
Dilithium2 & 43474 & 17472 & 18 & 3 & 11 & 294 & 5 & 8 \\
Dilithium3 & 42800 & 25010 & 36 & 6 & 22 & 654 & 10 & 20 \\
Dilithium5 & 44171 & 18591 & 12 & 8 & 14 & 355 & 4 & 13 \\
Falcon-512 & 60647 & 18641 & 18 & 16 & 50 & 106 & 4 & 16 \\
Falcon-1024 & 60794 & 9360 & 8 & 8 & 22 & 40 & 2 & 7 \\
Falcon-padded-512 & 60678 & 9396 & 8 & 9 & 21 & 41 & 2 & 6 \\
Falcon-padded-1024 & 61128 & 9456 & 8 & 6 & 19 & 42 & 2 & 9 \\
MAYO-1 & 43542 & 6924 & 8 & 3 & 4 & 33 & 2 & 8 \\
MAYO-2 & 43533 & 6984 & 8 & 4 & 5 & 32 & 2 & 8 \\
MAYO-3 & 46823 & 6996 & 8 & 3 & 4 & 35 & 2 & 6 \\
MAYO-5 & 44302 & 6528 & 6 & 4 & 3 & 34 & 3 & 11 \\
ML-DSA-44 & 43668 & 17616 & 0 & 3 & 1 & 21 & 1 & 5 \\
ML-DSA-44-ipd & 43338 & 8820 & 9 & 3 & 4 & 172 & 3 & 8 \\
SPHINCS+-SHA2-128f & 37109 & 8052 & 8 & 3 & 5 & 112 & 6 & 9 \\
SPHINCS+-SHA2-128s & 37379 & 8052 & 8 & 3 & 8 & 109 & 5 & 7 \\
SPHINCS+-SHA2-192f & 42143 & 8460 & 7 & 3 & 9 & 113 & 5 & 7 \\
SPHINCS+-SHA2-192s & 42535 & 8484 & 7 & 3 & 10 & 109 & 4 & 6 \\
SPHINCS+-SHA2-256f & 42437 & 8532 & 8 & 3 & 4 & 94 & 3 & 8 \\
SPHINCS+-SHA2-256s & 42758 & 8556 & 8 & 3 & 6 & 98 & 3 & 7 \\
cross-rsdp-128-balanced & 44024 & 10032 & 9 & 4 & 4 & 209 & 2 & 8 \\
\bottomrule
\end{tabular}
\caption{Results from scans on the \texttt{liboqs} library, showing various issues encountered during signature operations for each digital signature scheme, along with the total number of assembly executions and candidate gadgets.}
\label{tab:pqcdss}
\end{table*}

%Alternatively, the attacker can target the signature generation routine on the client's end to craft invalid signatures. The client cannot connect to any server by using the invalid signatures. Thus, it is possible to perform a denial of service (DoS) attack by breaking the signature generation routine. 
We find \attack gadgets on FIPS 204 standard, also on other PQC digital signatures schemes, FALCON~\cite{fouque2018falcon}, MAYO~\cite{beullens2021mayo}, and CROSS~\cite{baldi2023cross}.

Table \ref{tab:pqcdss} summarizes \attack gadgets found in the \texttt{liboqs} library for different PQC digital signature schemes. 
%ML-DSA, the implementation of the FIPS 204 standard, had fewer gadgets compared to the Dilithium reference implementation.
%Another interesting finding was related to the ML-DSA scheme, an implementation aligned with the FIPS 204 standard. 
Compared to Dilithium, ML-DSA, the implementation of the FIPS 204 standard, had fewer \attack gadgets, suggesting that its implementation might be more resilient to the specific fault attacks we conducted. However, this does not imply immunity, as the gadgets found were still capable of bypassing critical functions. The relatively lower number of vulnerabilities in ML-DSA could also be attributed to its simpler structure, which reduces the surface area for potential control flow subversion attacks.

We also evaluated SPHINCS+, a hash-based signature scheme standardized in FIPS 205. SPHINCS+ offers a different security foundation, relying on the hardness of hash-based constructions rather than lattice problems. While this scheme is robust against certain classes of attacks, our analysis uncovered several LeapFrog gadgets capable of bypassing signature verification. This suggests that even though the algorithm itself is designed to withstand quantum and classical cryptanalytic attacks, practical vulnerabilities arise due to implementation flaws that allow Rowhammer-based attacks to alter execution paths. Interestingly, the number of \attack gadgets identified in SPHINCS+ varied significantly based on its parameter set, with some configurations being more resilient than others. This highlights the importance of parameter selection in mitigating the risk of physical attacks.

\begin{comment}
\begin{table}
\centering
\begin{tabularx}{\linewidth}{c c c}
\toprule
Digital Signature Scheme & \makecell{Valid Sig.\\ Not Verified\\ (\# Gadgets)} & \makecell{Invalid Sig.\\ Incorrectly Verified\\ (\# Gadgets)} \\
\midrule
Dilithium2 & 294 & 5\\
Dilithium3 & 654 & 10\\
Dilithium5 & 355 & 4\\
Falcon-512 & 106 & 4\\
Falcon-1024 & 40 & 2\\
Falcon-padded-512 & 41 & 2\\
Falcon-padded-1024 & 42 & 2\\
MAYO-1 & 33 & 2\\
MAYO-2 & 32 & 2\\
MAYO-3 & 35 & 2\\
MAYO-5 & 34 & 3\\
ML-DSA-44 & 21 & 1\\
ML-DSA-44-ipd & 172 & 3\\
%ML-DSA-44-ipd copy & 172 & 3\\
SPHINCS+-SHA2-128f-simple & 112 & 6\\
SPHINCS+-SHA2-128s-simple & 109 & 5\\
SPHINCS+-SHA2-192f-simple & 113 & 5\\
SPHINCS+-SHA2-192s-simple & 109 & 4\\
SPHINCS+-SHA2-256f-simple & 94 & 3\\
SPHINCS+-SHA2-256s-simple & 98 & 3\\
cross-rsdp-128-balanced & 209 & 2\\
\bottomrule
\end{tabularx}
\caption{Results from scans on the \texttt{liboqs} library, showing the number of LeapFrog gadgets found in the code that would result in either breaking the scheme's ability to verify a signature, or cause a signature scheme to verify an invalid signature falsely.}
\label{tab:pqcdss}
\end{table}
\end{comment}

%The FIPS 205 standard is not implemented in the \texttt{liboqs} library. However, we can scan SPHINCS+~\cite{aumasson2019sphincs}, which is the selected algorithm for FIPS 205. 
SPHINCS+ has more gadgets compared to the FALCON-1024 configuration, but in some configurations, it has fewer gadgets than FALCON-512, another selected algorithm that is not standardized. Overall, our findings indicate that there are generally more \attack gadgets that enable bypassing signature verification compared to those that can falsely verify an invalid signature, indicating higher feasibility for DoS attacks with lower security impact compared to impersonation attacks.

\subsection{TLS Handshake}\label{sec:tls_attack}

\begin{table}[]
    \centering
    \begin{tabular}{c|c |c| c| c| c}
    \toprule
        Target & Size & \#Inst.\textsubscript{exec}& $d_{HD}$ & \multicolumn{2}{c}{\# Candidates} \\
          &&& &  \blackcircled{2} on & \blackcircled{2} off\\
    \midrule
          \multirow{3}{*}{TLS} &\multirow{3}{*}{29KB}  &\multirow{3}{*}{5328007}&  1 & 315 & 2493\\
                        &&&  2 & 2240 & 14413\\
                        &&&  3 & 21841 & 67421 \\
    % \midrule
    %       \multirow{3}{*}{OpenSSL} &\multirow{3}{*}{818KB} &\multirow{3}{*}{49431} &  1 & N/A & 2700  \\
    %                     &&&  2 & N/A & 20208 \\
    %                     &&&  3 & N/A & 70475 \\

    % \midrule
    %       \multirow{3}{*}{sudo} &\multirow{3}{*}{227KB}  &\multirow{3}{*}{148177}&  1 & 1100  & 8655     \\
    %                     &&&  2 & 6910  & 54203    \\ % 6.2 seconds 13.5 sec
    %                     &&&  3 & 30908 &  221181  \\ % 2.3 seconds 14.3 sec
    %  \midrule
    %       \multirow{3}{*}{Dilithium} &\multirow{3}{*}{84KB} &\multirow{3}{*}{24762} &  1 & N/A & 799  \\
    %                     &&&  2 & N/A & 4899 \\
    %                     &&&  3 & N/A & 18529 \\
                    
    \bottomrule
    \end{tabular}
    \caption{Number of gadget candidates found in TLS scenario for fault models with different Hamming distances. %We ran OpenSSL with \texttt{aria-128-cbc} cipher.
    }
    \label{tab:numberofgadgets}
\end{table}

In a full end-to-end attack example, we illustrate the potency of the attack by applying it within a client/server authentication framework, specifically using OpenSSL for signature verification.  Here, we consider a scenario where the attacker shares a physical computing space with the client. The goal of the attacker is to manipulate the client's signature verification mechanism, causing it to erroneously validate a corrupted signature as genuine. This manipulation forms part of a broader man-in-the-middle strategy, aimed at deceiving the client into believing they are securely connected to the intended server. Note that this attack also successfully achieves synchronization as part of the end-to-end demonstration due to the large attack window enabled by a network connection.

In the standard communication flow, the client initiates contact with the server by dispatching a \texttt{ClientHello} message. The server replies with a \texttt{ServerHello} message, which carries its public key and a digital signature of the handshake process. The client's role is then to authenticate this signature using the server's public key. Under normal circumstances, a verified signature would indicate a secure channel, prompting the client to transmit sensitive data to the server. However, in our attack scenario, the attacker strategically alters the signature verification process at the client's end. By inducing a single-bit error during this process, the client is misled into accepting a fraudulent signature as valid. As a result, the client erroneously trusts the communication channel and proceeds to send sensitive information to the attacker.

% TODO: Change this diagram with actual (altered) attack diagram [DONE]

\begin{figure}
    \centering
    \includegraphics[width=0.9\columnwidth]{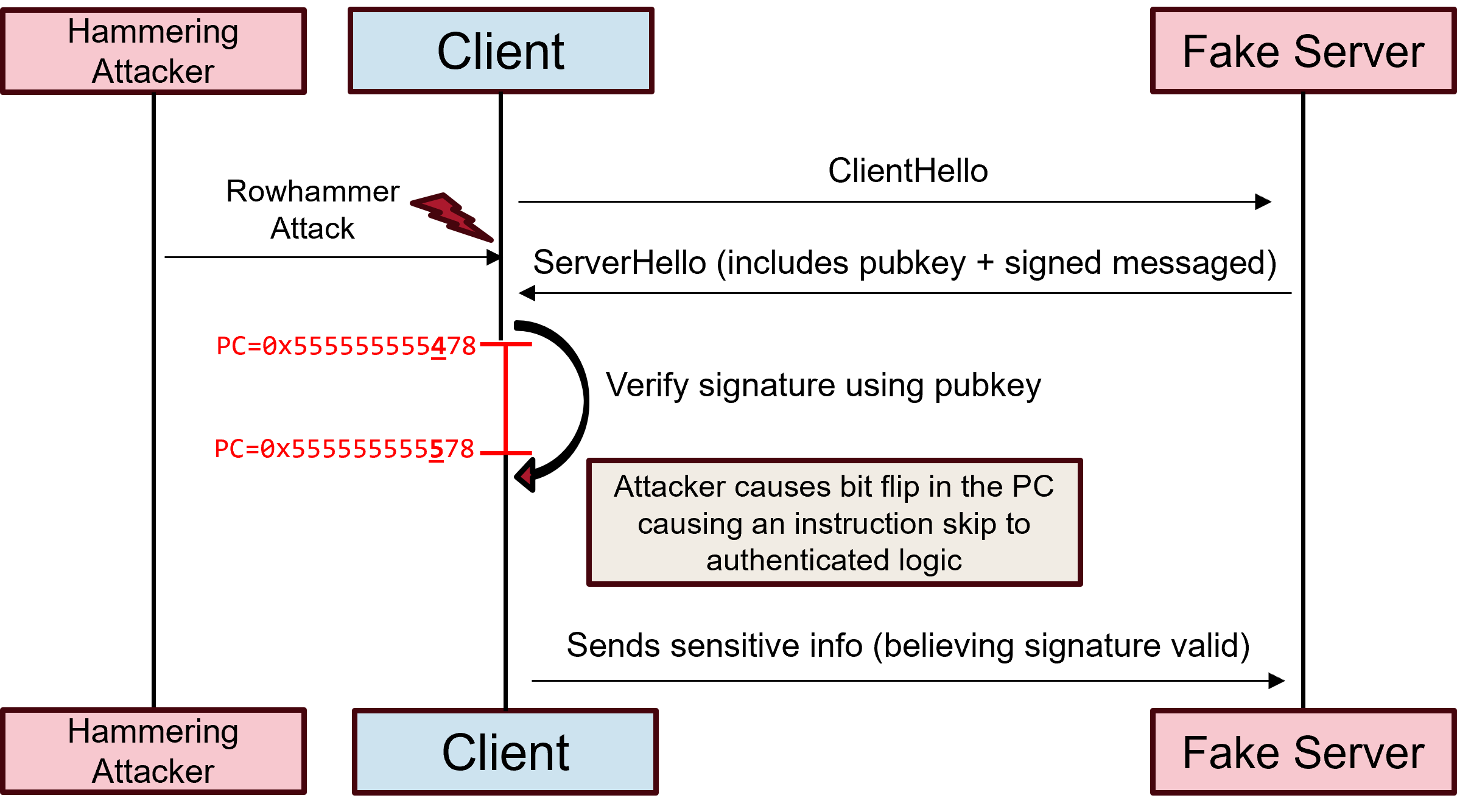}
    \caption{TLS Handshake: The client attempts to authenticate the server, and a colocated rowhammer attacker flips the PC value causing an instruction skip resulting in a misauthentication - this is an end-to-end attack}
    \label{fig:poc1}
\end{figure}

Figure \ref{fig:poc1} illustrates a standard interaction where the client establishes a connection with the server, sends a request, and then receives a server-signed message, enabling server authentication. A critical aspect to note is the client's susceptibility to a Rowhammer attack while it awaits the server's response. This waiting period, which can last several milliseconds, is primarily dictated by the server's response time. During this interval, an attacker has the opportunity to exploit the Rowhammer vulnerability by targeting the client's memory.

\begin{figure}[h]
    \centering
    \includegraphics[width=\columnwidth]{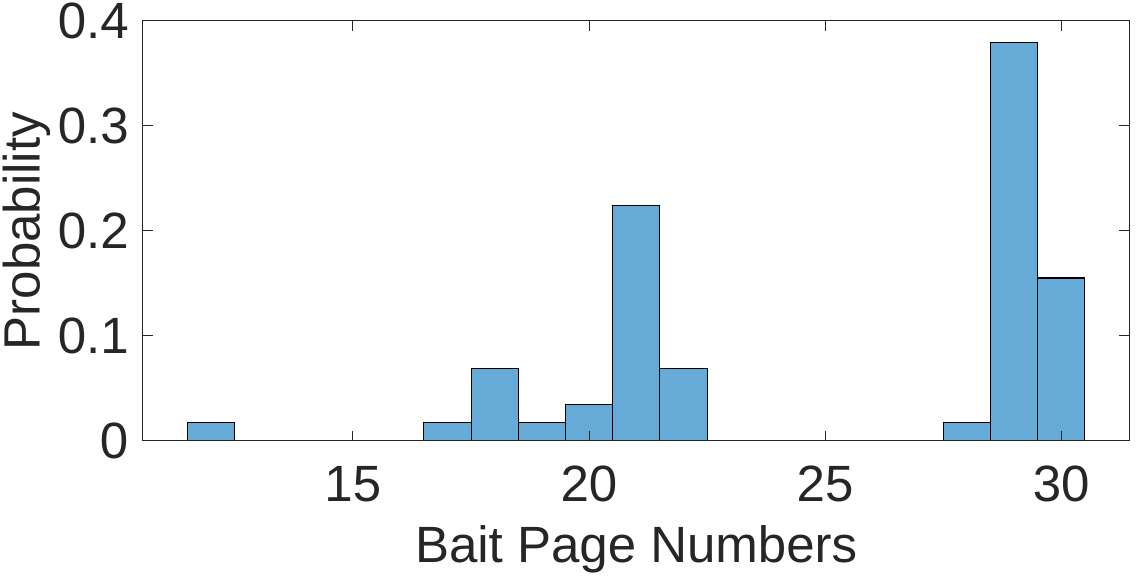}
    \caption{Probability distribution of bait page numbers.}
    \label{fig:baitpages}
\end{figure}

We first profile the target memory to determine where the client places the PC value in recently deallocated pages. We use a process we call \textit{baiting}~\cite{flipfengshui} where we allocate a series of pages as the attacker. Then, we deallocate the flippy page, forcing the victim process to use recently-released flippy locations, where the PC values are stored. You can see our results in Figure \ref{fig:baitpages} where we see that releasing 29 pages results in nearly a 30\% chance of a PC value being stored in the next page released (which would be the flippy page). 

% Results from scanning PoC

We scan the client binary using \tool while the server is using the correct and incorrect private key. With step \blackcircled{2} on, we found  315 unique gadget candidates with $d_{H}=1$. When the server uses the correct key, the client binary terminates with return code \texttt{0}, and when the server uses an incorrect key, the client returns with code \texttt{1}. In step \blackcircled{5}, we look for PC corruptions that cause the client to return with value \texttt{0}, meaning it incorrectly authenticates the server. After the simulation, we found that one of the candidates was a LeapFrog gadget that caused false authentication of the malicious server with an incorrect key. % Indeed, the found gadget was the one we created. 478-> 578. 

% TODO: Results on tool [DONE]
% Step 6 - Time simulation results here VVVV

%%%%%%%%%%%%%%%%%%%%%%%%%%%%

Then, we scan the client with step \blackcircled{2} off. With this mode, \tool detected 2493 gadget candidates with $d_{H}=1$. After the simulation steps, we verified that 21 of those candidates were LeapFrog gadgets that caused the client to return with \texttt{0}, including the one found earlier. The number of candidates for different Hamming distance values is given in Table~\ref{tab:numberofgadgets}.

The total time for the end-to-end attack to induce a successful misauthentication of the TLS handshake was 12 hours and 25 minutes, as seen in Table ~\ref{tab:all_results_openssl}. This time included profiling the system for the proper flippy pages with the correct offset, meaning the actual online time was around 2 hours. The experiment found a total of 1647 unique flippy pages, and over the course of the 2 hours of online attacking, we saw 2206 attacks out of ~7000 (given our 30\% bait page success rate) where the program counter was baited into the correct page we were attacking before it flipped. Note, this was not all one connection, but many handshakes - thus an interrupt would only affect one trial.

\begin{table}[h]
    \centering
    %\normalsize  % This ensures the table uses the normal font size
    \begin{tabular}{l|c c}
    \toprule
        Category & Result \\
    \midrule
        Total Time & 12 hrs 25 mins \\ 
        Online Time & 1 hr 54 mins \\ 
        Total Flippy Pages & 1647 \\ 
        Total Attacks w/ Correct \# of Bait pages & 2206 \\ 
    \bottomrule
    \end{tabular}
    \caption{Results from the end-to-end attack on code using OpenSSL client/server signature verification with LeapFrog gadget}
    \label{tab:all_results_openssl}
%\vspace{-1cm}
\end{table}

\section{Countermeasures}

\paragraph{Rowhammer Resistant Hardware}
Increasing the DRAM refresh rate is a commonly cited countermeasure to prevent Rowhammer attacks. Standard DRAM refresh is 64ms, meaning that a Rowhammer attack has 64ms to flip a bit before the row refreshes. Thus, a faster refresh rate will result in a shorter time window for the Rowhammer attack to be performed and should result in fewer flips. This is not an ideal solution, however, because a faster refresh rate will lead to worse power usage and performance overall. Alternative methods such as probabilistic row refresh \cite{Wang2021} and parallel row refresh \cite{HiRA2022} are not available in consumer systems. Additionally, upgrading hardware to newer DDR5 technology has also proven to be ineffective \cite{298232}. 

A novel countermeasure against Rowhammer attacks is the Randomized Row-Swap (RRS) method \cite{saileshwar2022randomized}. This approach fundamentally disrupts the spatial connection between aggressor and victim DRAM rows, thereby offering a robust defense against complex Rowhammer access patterns, including those not mitigated by victim-focused methods like the Half-Double attack. RRS operates by periodically swapping aggressor rows with randomly selected rows within the DRAM memory, limiting the potential damage to any single locality. While RRS can be implemented in conjunction with any tracking mechanism, its effectiveness has been demonstrated when paired with a Misra-Gries tracker, targeting a Row Hammer Threshold of 4.8K activations, akin to state-of-the-art attacks. 

Initial beliefs held that the Error Correcting Code (ECC) would serve as an effective defense against Rowhammer attacks. However, subsequent research has shown that ECC, despite its prevalence in server environments, falls short of a comprehensive solution. This inadequacy primarily arises due to ECC's vulnerability to scenarios involving triple bit flips, a phenomenon well-documented in the literature \cite{cojocar2019ecc}. Additionally, ECC, while standard in server-grade hardware, is typically absent in consumer-grade DRAM systems.

\paragraph{Adding \texttt{nops} To Code}
A mitigation against the LeapFrog attack specifically would be patching the source code or binary such that it is no longer vulnerable. Given the single-bit flip requirement of Rowhammer on the PC values, adding enough \texttt{nop}s within the \attack gadget to prevent instruction skips that only require a single-bit flip would potentially mitigate the attack. Adding \texttt{nop} instructions to source code is not trivial when the compiler optimizations are enabled since the compiler may reorder the critical parts in a different way, which makes the patch ineffective.
A mitigation tool that adds \texttt{nop}s to binary itself may overcome the compiler effect. Yet, adding new instructions to a binary will result in a change in the address of all the following instructions, which may introduce new \attack gadgets. Therefore, the patched binary needs to be re-evaluated if the new version still has gadgets. Although a LeapFrog-aware compiler may potentially generate a \attack proof binary, we claim it is not a sound and reliable approach. 

\paragraph{Adding Redundancy to the Control Flow}
Since LeapFrog gadgets are hard to mitigate in the source code and binary manually, we need a generic mitigation that can be implemented at the compiler level. The main target in LeapFrog is the program counter values that are temporarily pushed into the stack. Pushing multiple copies of the program counter to the stack and making sure the ultimate decision to return to an address is made on the combination of these copies would potentially make the attack impractical.

% \section{Discussion}

% \section{Disclosures}

% We have disclosed this vulnerability to the SUDO team. 

\section{Conclusion}
In this work, we introduced \attack, a specific type of Rowhammer exploit that directly targets the control flow of programs by manipulating the Program Counter stored in the stack. This novel approach marks a significant shift in the understanding of Rowhammer threats, moving beyond traditional data integrity attacks to those that can alter program execution. Our successful demonstration of this attack in an OpenSSL TLS handshake scenario highlights its practical effectiveness and potential impact on widely used security protocols.

Furthermore, we proposed a systematic approach to identify LeapFrog gadgets in real-world software. Using our \tool analysis tool, we scanned multiple OpenSSL ciphers, Open Quantum Safe signature schemes, and machine learning classification algorithms and quantified the number of LeapFrog gadgets in this software. Even though the identification of vulnerable software is relatively straightforward thanks to our detection tool, mitigation of \attack is not a trivial task since it is not transparent to the developers on a source code level. Instead, dedicated Rowhammer-resistant DRAM hardware or Rowhammer-aware compiler tools will be required to prevent LeapFrog attacks.

\section{Open Science Statement}
The code used for \tool has been open-sourced and can be found here -  \url{https://github.com/andrewadiletta/leapfrog_scanner}

\section{Disclaimer}
Andrew Adiletta's affiliation with The MITRE Corporation is provided for identification purposes only, and is not intended to convey or imply MITRE's concurrence with, or support for, the positions, opinions, or viewpoints expressed by the author. All references are public domain.

% TODO: Discussion and Conclusion

% \section{Disclaimer}
% Andrew Adiletta's affiliation with The MITRE Corporation is provided for identification purposes only, and is not intended to convey or imply MITRE's concurrence with, or support for, the positions, opinions, or viewpoints expressed by the author.

% \section*{Acknowledgements}
% This work was supported by the National Science Foundation grant CNS-2026913 and in part by a grant from the Qatar National Research Fund.

\newpage

\bibliographystyle{plain}
\bibliography{references}

\end{document}